\documentclass[12pt,a4paper]{article}
\pdfoutput=1
\usepackage{epsfig}
\usepackage{amssymb}
\usepackage{graphicx}
\usepackage{color}
\usepackage{subfigure}
\usepackage{mathtools}
\usepackage[hidelinks]{hyperref}
\makeatletter
\renewcommand{\theequation}{\thesection.\arabic{equation}}
\@addtoreset{equation}{section} \makeatother
\renewcommand{\vec}[1]{\underline{#1}}

\newcommand{\sss}[1]{\scriptscriptstyle{#1}}
\newcommand{\lra}[1]{\la{#1}\ra}

\def\a{\alpha}
\def\b{\beta}

\def\ep{\epsilon}
\def\e{\eta}
\def\ph{\phi}
\def\Ph{\Phi}

\def\l{\lambda}

\def\m{\mu}

\def\th{\theta}

\def\s{\sigma}
\def\S{\Sigma}
\def\ta{\tau}
\def\o{\omega}

\def\pr{\prime}
\def\z{\zeta}

\def\lt{\left}
\def\rt{\right}

\def\nn{\nonumber}

\DeclareMathOperator{\tr}{Tr}

\def\p{\partial}

\def\la{\langle}
\def\ra{\rangle}

\def\mca{\mathcal{A}}
\def\mcb{\mathcal{B}}
\def\mcs{\mathcal{S}}
\def\mcm{\mathcal{M}}

\def\nn{\nonumber}
\def\bul{\noindent$\bullet$}

\def\bea{\begin{eqnarray}}
\def\eea{\end{eqnarray}}

\begin{document}

\begin{titlepage}
\title{\vskip -60pt
\vskip 20pt %
Junctions of mass-deformed nonlinear sigma models on $SO(2N)/U(N)$ and $Sp(N)/U(N)$ I
}
\author{
Taegyu Kim\footnote{e-mail : taegyukim@skku.edu}~ and 
Sunyoung Shin\footnote{e-mail : sihnsy@skku.edu}}
\date{}
\maketitle \vspace{-1.0cm}
\begin{center}
~~~
\it  Department of Physics and Institute of Basic Science, \\Sungkyunkwan University, Suwon 16419, Republic of Korea

%~~~\\
%~~~\\
\end{center}

\thispagestyle{empty}

\begin{abstract}
We construct on-shell ${\mathcal{N}}=2$ nonlinear sigma models on $SO(2N)/U(N)$ and $Sp(N)/U(N)$ by holomorphically embedding the models in the hyper-K\"{a}hler nonlinear sigma model on the cotangent bundle of the Grassmann manifold $T^\ast G_{2N,N}$ in the ${\mathcal{N}}=1$ superspace formalism. We apply the moduli matrix formalism to the mass-deformed nonlinear sigma models on the quadrics to study three-pronged junctions by using a recently proposed diagram method.
\end{abstract}

%\pacs{Valid PACS appear here}% PACS, the Physics and Astronomy Classification Scheme.
%{\small
%\begin{flushleft}
%%%
%~~~~~~~~\textit{PACS}: \\
%~~~~~~~~\textit{Keywords}:
%\end{flushleft}}
%%%
%\thispagestyle{empty}
\end{titlepage}

\vfill
\newpage
\setcounter{page}{1}
\setcounter{footnote}{0}
\renewcommand{\thefootnote}{\arabic{footnote}}

%%%%%%%%%%%%%%%%%%%%%%%%%%%%%%%%%%%%%%%%%%%%%%%%%%%%%%%%%%%%%%%%%%%%%%%%%%%%
\section{Introduction} \label{sec:intro}
\setcounter{equation}{0}
%%%%%%%%%%%%%%%%%%%%%%%%%%%%%%%%%%%%%%%%%%%%%%%%%%%%%%%%%%%%%%%%%%%%%%%%%%%%

The number of supersymmetries of a nonlinear sigma model is related to the target space geometry. Rigid supersymmetric nonlinear sigma models with four supercharges and with eight supercharges are K\"{a}hler manifolds and hyper-K\"{a}hler manifolds respectively \cite{Zumino:1979et,AlvarezGaume:1981hm,Calabi,Hitchin:1986ea}. 

K\"{a}hler nonlinear sigma models on the Hermitian symmetric spaces are constructed as gauge theories \cite{Higashijima:1999ki,Higashijima:2001vk}. Hyper-K\"{a}hler nonlinear sigma models are constructed in the $\mathcal{N}=1$ superspace formalism \cite{Curtright:1979yz,AlvarezGaume:1980vs,Rocek:1980kc,Lindstrom:1983rt}. Hyper-K\"{a}hler nonlinear sigma models on the cotangent bundles of the complex projective space and the Grassmann manifold are constructed in the harmonic superspace formalism \cite{Galperin:1985dw,Galperin:2001uw}. Hyper-K\"{a}hler nonlinear sigma models on the cotangent bundles of the Hermitian symmetric spaces are constructed in the projective superspace formalism \cite{Arai:2006qt,Arai:2006gg,Arai:2007cx}. 

We are interested in nonlinear sigma models where the fields are homogeneous coordinates since the hyper-K\"{a}hler nonlinear sigma model on $T^\ast G_{N_F,N_C}$\footnote{$G_{N+M,M}=\frac{SU(N+M)}{SU(N)\times SU(M)\times U(1)}$.} is the strong coupling limit of the $U(N_C)$ gauge theory. The symmetric spaces $SO(2N)/U(N)$ and $Sp(N)/U(N)$ are quadrics in the Grassmann manifold $G_{2N,N}$ \cite{Hua}. The K\"{a}hler nonlinear sigma models on $SO(2N)/U(N)$ and $Sp(N)/U(N)$ are holomorphically embedded in the K\"{a}hler nonlinear sigma model on $G_{2N,N}$, where the fields are homogeneous coordinates, by the Lagrange multiplier method in the ${\mathcal{N}}=1$ superspace formalism. This amounts to introducing superpotentials into the nonlinear sigma model on the Grassmann manifold \cite{Higashijima:1999ki}. Therefore, the nonlinear sigma models on $SO(2N)/U(N)$ and $Sp(N)/U(N)$ are of interest since the models have superpotentials and the models are realised as non-Abelian gauge theories. In this paper, we construct on-shell $\mathcal{N}=2$ nonlinear sigma models on $SO(2N)/U(N)$ and $Sp(N)/U(N)$ by holomorphically embedding the models in the nonlinear sigma model on $T^\ast G_{2N,N}$ in the $\mathcal{N}=1$ superspace formalism by following the method of \cite{Higashijima:1999ki}.

Potential terms and Bogomol'nyi-Prasad-Sommerfield (BPS) objects are discussed in \cite{AlvarezGaume:1983ab,Abraham:1992vb,Gauntlett:2000bd,Gauntlett:2000ib}. The moduli matrix formalism was proposed to study walls of $\mathcal{N}=2$ non-Abelian gauge theories \cite{Isozumi:2004jc,Isozumi:2004va}. The moduli matrix formalism has been applied to various BPS objects \cite{Eto:2006pg,Eto:2005yh,Eto:2005cp,Eto:2005fm,Gudnason:2010rm,Eto:2011cv,Eto:2020vjm}. Intersecting walls form junctions \cite{Gauntlett:2000bd,Eto:2005cp,Eto:2005fm,Eto:2020vjm,Abraham:1990nz,Gibbons:1999np,Carroll:1999wr,Saffin:1999au,Nam:2000qe}. Three-pronged junctions are discussed in the moduli matrix formalism \cite{Eto:2005cp,Eto:2005fm}. In \cite{Eto:2005fm}, junctions of the mass-deformed nonlinear sigma model on the Grassmann manifold are studied in the moduli matrix formalism by embedding the Grassmann manifold into the complex projective space via the Pl\"{u}cker embedding.

A pictorial representation was proposed to elaborate the moduli matrix formalism \cite{Lee:2017kaj,Arai:2018tkf}. An alternative method is proposed in \cite{Shin:2018chr} to construct three-pronged junctions of the mass-deformed nonlinear sigma model on the Grassmann manifold. The moduli matrix formalism is directly applied to three-pronged junctions of the mass-deformed nonlinear sigma model on the Grassmann manifold by making use of diagrams \cite{Eto:2005cp,Shin:2018chr} in the pictorial representation \cite{Lee:2017kaj,Arai:2018tkf}, since the Pl\"{u}cker embedding can be avoided in this approach. 

%It is expected that we can construct three-pronged junctions of the on-shell $\mathcal{N}=2$ nonlinear sigma models on $SO(2N)/U(N)$ and $Sp(N)/U(N)$ that will be proposed in this paper by using the diagram method \cite{Eto:2005cp,Shin:2018chr}, as $SO(2N)/U(N)$ and $Sp(N)/U(N)$ are quadrics in the Grassmann manifold, and the Pl\"{u}cker embedding is not required in the pictorial representation.

The quadrics $SO(2N)/U(N)$ and $Sp(N)/U(N)$ are submanifolds of the Grassmann manifold. Therefore, it is expected that we can construct three-pronged junctions of the on-shell $\mathcal{N}=2$ mass-deformed nonlinear sigma models on $SO(2N)/U(N)$ and $Sp(N)/U(N)$ that will be proposed in this paper by using the moduli matrix formalism \cite{Isozumi:2004jc,Isozumi:2004va} and the diagram method \cite{Eto:2005cp,Shin:2018chr} as the Pl\"{u}cker embedding is not required in the pictorial representation that is proposed in \cite{Lee:2017kaj,Arai:2018tkf}.

The purpose of this paper is to construct on-shell ${\mathcal{N}}=2$ nonlinear sigma models on $SO(2N)/U(N)$ and $Sp(N)/U(N)$ in the ${\mathcal{N}}=1$ superspace formalism and to study three-pronged junctions of the mass-deformed nonlinear sigma models on the quadrics by using the moduli matrix formalism and the diagram method.

This paper is organised as follows. In Section \ref{sec:model}, we construct on-shell ${\mathcal{N}}=2$ mass-deformed nonlinear sigma models on $SO(2N)/U(N)$ and $Sp(N)/U(N)$ in the ${\mathcal{N}}=1$ superspace formalism. In Section \ref{sec:bps_sol}, we discuss mass-deformed nonlinear sigma models on $SO(2N)/U(N)$ and $Sp(N)/U(N)$ with complex masses, which are derived from the ${\mathcal{N}}=2$ mass-deformed nonlinear sigma models obtained in Section \ref{sec:model}, and apply the moduli matrix formalism  to the BPS equations and the constraints. In Section \ref{sec:3pronged_junct}, we briefly review vacua, walls and junctions in the moduli matrix formalism and study three-pronged junctions of the mass-deformed nonlinear sigma models on $SO(8)/U(4)$ and $Sp(3)/U(3)$ by using the diagram method. In Section \ref{sec:summary}, we summarise our results.

%%%%%%%%%%%%%%%%%%%%%%%%%%%%%%%%%%%%%%%%%%%%%%%%%%%%%%%%%%%%%%%%%%%%%%%%%%%
\section{${\mathcal{N}}=2$ nonlinear sigma models on $SO(2N)/U(N)$ and $Sp(N)/U(N)$ in the $\mathcal{N}=1$ superspace formalism} 
\label{sec:model}
\setcounter{equation}{0}
%%%%%%%%%%%%%%%%%%%%%%%%%%%%%%%%%%%%%%%%%%%%%%%%%%%%%%%%%%%%%%%%%%%%%%%%%%%%
%
In this section, we construct on-shell ${\mathcal{N}}=2$ nonlinear sigma models on $SO(2N)/U(N)$ and $Sp(N)/U(N)$ by holomorphically embedding the models in the hyper-K\"{a}hler nonlinear sigma model on the cotangent bundle of the Grassmann manifold $T^\ast G_{2N,N}$ in the $\mathcal{N}=1$ superspace formalism. In Section \ref{subsec:grssmodel}, we review the hyper-K\"{a}hler nonlinear sigma model on the cotangent bundle of the Grassmann manifold $T^\ast G_{N+M,M}$ in the $\mathcal{N}=1$ superspace formalism \cite{Lindstrom:1983rt,Arai:2002xa,Arai:2003es,Arai:2003tc}, which has on-shell ${\mathcal{N}}=2$ supersymmetry. In Section \ref{subsec:sospmodel}, we construct on-shell ${\mathcal{N}}=2$ nonlinear sigma models on $SO(2N)/U(N)$ and $Sp(N)/U(N)$ in the $\mathcal{N}=1$ superspace formalism.

%%%%%%%%%%%%%%%%%%%%%%%%%%%%%%%%%%%%%%%%%%%%%%%%%%%%%%%%%%%%%%%%%%%%%%%%%%%
\subsection{Hyper-K\"{a}hler nonlinear sigma model on the cotangent bundle of the Grassmann manifold $T^\ast G_{N+M,M}$ }  
\label{subsec:grssmodel}
\setcounter{equation}{0}
%%%%%%%%%%%%%%%%%%%%%%%%%%%%%%%%%%%%%%%%%%%%%%%%%%%%%%%%%%%%%%%%%%%%%%%%%%%%
%
The hyper-K\"{a}hler nonlinear sigma model on $T^\ast G_{N+M,M}$ is discussed in the $\mathcal{N}=1$ superspace formalism in \cite{Lindstrom:1983rt,Arai:2002xa,Arai:2003es,Arai:2003tc}:
\begin{align}\label{eq:ordnryfld_act}
S=\int d^4x \Big\{&\int d^4\th \tr\Big(\Ph\bar{\Ph} e^V+\bar{\Psi}\Psi e^{-V}-c^\prime V \Big) \nn \\
&+ \Big[ \int d^2\th \tr\Big(\Xi\lt(\Ph\Psi-b I_M \rt)\Big)+\mathrm{(conjugate~transpose)}\Big]\Big\} ,\nn\\
&~(c^\prime \in \mathbf{R}_{\neq 0},~ b\in \mathbf{C}). 
\end{align}
The nonlinear sigma model (\ref{eq:ordnryfld_act}) has on-shell $\mathcal{N}=2$ supersymmetry.  The ${\mathcal{N}}=2$ hypermultiplet consists of ${\mathcal{N}}=1$ chiral field $\Phi$ and ${\mathcal{N}}=1$ chiral field $\Psi$. The ${\mathcal{N}}=2$ vector multiplet consists of ${\mathcal{N}}=1$ vector field $V$ and ${\mathcal{N}}=1$ chiral field $\Xi$. Chiral field $\Phi$ is an $M\times (N+M)$ matrix, chiral field $\Psi$ is an $(N+M)\times M$ matrix, vector field $V$ is an $M\times M$ matrix and complex field $\Xi$ is an $M\times M$ matrix. We diagonalise $\Xi$ for later use. We follow the convention of \cite{Galperin:2001uw}. The constants $b$, $b^\ast$ and $c^\prime$ are the Fayet-Iliopoulos (FI) parameters. 

The Lagrangian has constraints 
\begin{align}
&\Ph\bar{\Ph} e^V-e^{-V}\bar{\Psi}\Psi -c^\prime I_M=0, \label{eq:cons1_ordnryfld} \\
&\Ph\Psi-bI_M=0,\quad \mbox{(c.t.)}=0.  \label{eq:cons2_ordnryfld}
\end{align}
%
%The vector field $V$ can be solved. We focus on the case where $c^\prime=c>0$. The K\"{a}hler potential for the Lindstr\"{o}m-Roc\v{e}k metric is \cite{Lindstrom:1983rt,Arai:%2003tc}
%%
%\begin{align} \label{eq:ordfldhkpot}
%K=&\tr \sqrt{c^2I_M+4\Ph\bar{\Ph}\bar{\Psi}\Psi}-c\tr\ln\Big(cI_M+\sqrt{c^2I_M+ 4\Ph\bar{\Ph}\bar{\Psi}\Psi}\Big) \nn\\
%&+c\tr\ln(\Ph\bar{\Ph}).
%\end{align}
%%
%%
The constraint (\ref{eq:cons2_ordnryfld}) can be solved by two cases, $b=0$ and $b\neq0$ with proper gauge fixing. The two cases are related by an $SU(2)_R$ transformation, which does not preserve the holomorphy \cite{Lindstrom:1983rt,Arai:2002xa,Arai:2003tc}. The parametrisation, which is considered in \cite{Arai:2003tc}
is \\

\bul $b=0$, $c^\prime=c>0$
\bea \label{eq:b=0}
&\Ph=\lt(I_M~~f \rt),~~
\Psi=\lt(\begin{array}{c}
-f g \\
g
\end{array}\rt).
\eea

\bul $b\neq 0$ \cite{AlvarezGaume:1980vs,Rocek:1980kc,Lindstrom:1983rt}
\bea  \label{eq:bneq0}
\Ph=Q\lt(I_M ~~ s\rt),~
\Psi=\lt(\begin{array}{c}
I_M \\ t
\end{array}
\rt)Q,~~Q=\sqrt{b}\lt(I_M+ s t\rt)^{-\frac{1}{2}}.
\eea
Fields $f$ and $s$ are $M\times N$ matrices. Fields $g$ and $t$ are $N\times M$ matrices\footnote{
For the case of $b=0$, $c^\prime < 0$, instead of (\ref{eq:b=0}), the constraint $\Phi\Psi=0$ can be solved by the following parametrisation: 
\bea \label{eq:b=0pr}
&\Phi=\lt(-uv~~u \rt),~~
\Psi=\lt(\begin{array}{c}
I_M \\
v
\end{array}\rt).
\eea
Field $u$ is an $M \times N$ matrix and field $v$ is an $N \times M$ matrix.}. 

The K\"{a}hler potential for the Lindstr\"{o}m-Roc\v{e}k metric \cite{Lindstrom:1983rt,Arai:2003tc} is obtained by solving the vector field $V$. We are interested in the $b=0$ case in this paper. With the parametrisation (\ref{eq:b=0}), the potential is
\begin{align}\label{eq:compfldkhpot}
K=&\tr\sqrt{c^2I_M+4(I_M+f\bar{f})\bar{g}(I_N+\bar{f} f)g} \nn\\
&-c\tr\ln\lt(cI_M+\sqrt{c^2I_M+4(I_M+f\bar{f})\bar{g}(I_N+\bar{f} f)g}\rt) \nn\\
&+c\tr\ln(I_M+f\bar{f}).
\end{align}
For $g=0$, the potential (\ref{eq:compfldkhpot}) becomes the potential of the Grassmann manifold. Therefore, $f$ parametrises the base Grassmann manifold whereas $g$ parametrises the cotangent space as the fiber \cite{Arai:2003tc}. 

There is a bundle structure in the nonlinear sigma model (\ref{eq:ordnryfld_act}) with $b=0$ as we can see in (\ref{eq:compfldkhpot}). In the nonlinear sigma models with $b=0$, $c^\prime >0$ and the parametrisation (\ref{eq:b=0}), field $\Phi$ parametrises only the base Grassmann manifold \cite {Arai:2003tc}. The same argument holds true for the nonlinear sigma model with $b=0$, $c^\pr<0$ and the parametrisation is (\ref{eq:b=0pr}). In this case, field $\Psi$ parametrises only the base Grassmann manifold. Therefore, we can simply reduce the number of supersymmetry of the hyper-K\"{a}hler nonlinear sigma model in (\ref{eq:ordnryfld_act}) with $b=0$, $c^\pr\neq 0$ by setting $\Psi=0=\bar{\Psi}$($\Phi=0=\bar{\Phi}$) for $c^\prime >0$($c^\prime < 0$)\footnote{The parametrisation for $\Phi\Psi=0$ is
\begin{eqnarray}\label{eq:b=0prpr}
&\Ph=\lt(k_+~~i\sqrt{k_+ k_-} \rt),~~
\Psi=\lt(\begin{array}{c}
k_- \\
i\sqrt{k_+ k_-}
\end{array}\rt),
\end{eqnarray} in \cite{Lindstrom:1983rt}. In this case we cannot observe the structure.}. This is discussed in \cite{Isozumi:2004jc,Isozumi:2004va}\footnote{Interested readers may refer to Section III.F of \cite{Isozumi:2004va}. }. 

The mass-deformed nonlinear sigma model on $T^\ast G_{N+M,M}$ for $c^\prime =c >0$ \cite{Arai:2002xa,Arai:2003tc} is
\begin{align}\label{ea:n=2nlsmord}
S=\int d^4x \Big\{&\int d^4\th \tr\Big(\Ph\bar{\Ph} e^V+\bar{\Psi}\Psi e^{-V}-c V\Big) \nn \\
&+\Big[ \int d^2\th \tr\Big(\Xi\lt(\Ph\Psi-b I_M \rt)+\Ph M\Psi\Big) + \mathrm{(c.t.)}\Big]\Big\} ,\nn\\
&~~(c\in \mathbf{R}_{\geq0},~ b\in \mathbf{C}).
\end{align}
The mass-deformed nonlinear sigma model on $T^\ast G_{N+M,M}$ (\ref{ea:n=2nlsmord}) has on-shell ${\mathcal{N}}=2$ supersymmetry.  The relation between the component field action of (\ref{ea:n=2nlsmord}) and the component field action of the mass-deformed nonlinear sigma model on $T^\ast G_{N+M,M}$ that is constructed in the harmonic superspace formalism \cite{Galperin:2001uw} is presented in \ref{app:eq}. It is discussed in detail in \cite{Kim:2020obf}.

The relation between the bosonic component field action of the mass-deformed nonlinear sigma model on $T^\ast {\mathbf{C}}P^1$ in the $O(2)$ gauge invariant form that is constructed in the ${\mathcal{N}}=1$ superspace formalism and the bosonic component field action of the mass-deformed nonlinear sigma model that is constructed in the harmonic superspace formalism \cite{Galperin:1985dw,Galperin:2001uw} is identified in \cite{Arai:2003es}. The relation between the bosonic component field action of the mass-deformed nonlinear sigma model on $T^\ast G_{N,M}$ that is constructed in the ${\mathcal{N}}=1$ superspace formalism \cite{Lindstrom:1983rt} and the bosonic component field action of the mass-deformed nonlinear sigma model that is constructed in the harmonic superspace formalism \cite{Galperin:2001uw} is identified in \cite{Arai:2003tc}.

%%%%%%%%%%%%%%%%%%%%%%%%%%%%%%%%%%%%%%%%%%%%%%%%%%%%%%%%%%%%%%%%%%%%%%%%%%%
\subsection{On-shell $\mathcal{N}=2$ nonlinear sigma models on $SO(2N)/U(N)$ and $Sp(N)/U(N)$ }\label{subsec:sospmodel}
%\setcounter{equation}{0}
%%%%%%%%%%%%%%%%%%%%%%%%%%%%%%%%%%%%%%%%%%%%%%%%%%%%%%%%%%%%%%%%%%%%%%%%%%%%

The ${\mathcal{N}}=1$ nonlinear sigma models on $SO(2N)/U(N)$ and $Sp(N)/U(N)$ are constructed by holomorphically embedding the models in the K\"{a}hler nonlinear sigma model on the Grassmann manifold $G_{2N,N}$ \cite{Higashijima:1999ki}. The F-term constraint and its conjugate transpose are imposed by the Lagrange multiplier method. The invariant tensor $J$ is $\s^1 \otimes I_N$ for $SO(2N)/U(N)$ and $i\s^2 \otimes I_N$ for $Sp(N)/U(N)$. Vanishing F-term constraints are imposed so that the symmetries of the nonlinear sigma models are consistent with the gauge symmetry and the supersymmetry of the K\"{a}hler nonlinear sigma model on $G_{2N,N}$. It is shown that the nonlinear sigma models on the quadrics that are isomorphic to the complex projective spaces produce equivalent K\"{a}hler potentials \cite{Higashijima:2001vk}. It justifies the method and the results of \cite{Higashijima:1999ki}.

In this subsection, we construct on-shell ${\mathcal{N}}=2$ nonlinear sigma models on $SO(2N)/U(N)$ and $Sp(N)/U(N)$ by holomorphically embedding the models in the hyper-K\"{a}hler nonlinear sigma model on the cotangent bundle of the Grassmann manifold $T^\ast G_{2N,N}$ with the bundle structure that is discussed in Section \ref{subsec:grssmodel}. The hyper-K\"{a}hler nonlinear sigma model on the cotangent bundle of the Grassmann manifold $T^\ast G_{2N,N}$ (\ref{eq:ordnryfld_act}) with $b=0$ is
\begin{align}\label{eq:ordnryfld_act2}
S=\int d^4x \Big\{&\int d^4\th \tr\Big(\Ph\bar{\Ph} e^V+\bar{\Psi}\Psi e^{-V}-c^\prime V \Big) \nn \\
&+\Big[ \int d^2\th \tr\Big(\Xi\Ph\Psi\Big)+\mathrm{(conjugate~transpose)}\Big]\Big\} , \quad (c^\prime\in \mathbf{R}_{\neq0}).
\end{align}
The value of the FI parameter $c^\prime$ can be either positive or negative.

As we have seen in Section \ref{subsec:grssmodel} and \ref{app:eq}, the nonlinear sigma model (\ref{eq:ordnryfld_act2}) has on-shell ${\mathcal{N}}=2$ supersymmetry. The ${\mathcal{N}}=2$ hypermultiplet consists of ${\mathcal{N}}=1$ chiral field $\Phi$ and ${\mathcal{N}}=1$ chiral field $\Psi$. The ${\mathcal{N}}=2$ vector multiplet consists of ${\mathcal{N}}=1$ vector field $V$ and ${\mathcal{N}}=1$ chiral field $\Xi$. The constraint $\Phi\Psi=0$ can be solved by (\ref{eq:b=0}) or (\ref{eq:b=0pr}). In either case one of the chiral fields of the hypermultiplet parametrises only the base manifold. Therefore, regardless of the ${\mathcal{N}}=2$ vector multiplet, by setting one of the chiral fields of the hypermultiplet equal to zero, the ${\mathcal{N}}=2$ nonlinear sigma model on $T^\ast G_{2N,N}$ (\ref{eq:ordnryfld_act2}) should get reduced to the K\"{a}hler nonlinear sigma model on $G_{2N,N}$ since the nonlinear sigma model (\ref{eq:ordnryfld_act2}) is in the ${\mathcal{N}}=1$ superspace formalism.

We construct on-shell ${\mathcal{N}}=2$ nonlinear sigma models on $SO(2N)/U(N)$ and $Sp(N)/U(N)$ by holomorphically embedding the models in the hyper-K\"{a}hler nonlinear sigma model on the cotangent bundle of the Grassmann manifold $T^\ast G_{2N,N}$ (\ref{eq:ordnryfld_act2}), which is constrained by the relation $\Phi \Psi =0$. There should exist well-defined ${\mathcal{N}}=1$ nonlinear sigma models on $SO(2N)/U(N)$ and $Sp(N)/U(N)$ in the on-shell ${\mathcal{N}}=2$ nonlinear sigma models on $SO(2N)/U(N)$ and $Sp(N)/U(N)$ since the models are constructed in the ${\mathcal{N}}=1$ superspace formalism and the models have the bundle structure that is discussed in Section \ref{subsec:grssmodel}. 

The constraint $\Phi\Psi=0$ can be solved by (\ref{eq:b=0}) for $c^\pr > 0$ or by (\ref{eq:b=0pr}) for $c^\pr < 0$. With the parametrisation (\ref{eq:b=0}) for $c^\prime >0$, $\Phi$ parametrises only the base manifold. We impose the constraint $\Phi J \Phi^T=0$ and its conjugate transpose, which are the constraints of the ${\mathcal{N}}=1$ nonlinear sigma models on $SO(2N)/U(N)$ and $Sp(N)/U(N)$ \cite{Higashijima:1999ki}, on the base manifold as the models should have the ${\mathcal{N}}=1$ nonlinear sigma models on $SO(2N)/U(N)$ and $Sp(N)/U(N)$. The constraint $\Phi J \Phi^T=0$ restricts the other chiral field of the hypermultiplet to $\Psi^T J \Psi=0$\footnote{\bea \label{eq:fg_constraints} \Phi J \Phi^T = f^T +\ep f =0,~ \Psi^T J \Psi=-g^T (f^T+\ep f)g=0.\eea}. Therefore there exists the Lagrangian subject to the constraints $\Phi J \Phi^T=0$, $\Psi^T J \Psi=0$ and their conjugate transpose. 
With the parametrisation (\ref{eq:b=0pr}) for $c^\prime <0$, $\Psi$ parametrises only the base manifold. We impose the constraint $\Psi^T J \Psi=0$ on the base manifold. The constraint restricts the other chiral field to $\Phi J \Phi^T=0$. There exists the Lagrangian subject to the constraints $\Psi^T J \Psi=0$, $\Phi J \Phi^T=0$ and their conjugate transpose. Therefore, both of the constraints and their conjugate transpose should be imposed to obtain consistent on-shell ${\mathcal{N}}=2$ nonlinear sigma models on $SO(2N)/U(N)$ and $Sp(N)/U(N)$ in the ${\mathcal{N}}=1$ superspace formalism\footnote{This argument can also be justified by the fact that there are other parametrisations that solve the constraint $\Phi\Psi=0$, such as (\ref{eq:b=0prpr}).}.

The on-shell ${\mathcal{N}}=2$ nonlinear sigma models on $SO(2N)/U(N)$ and $Sp(N)/U(N)$ is
\begin{align}\label{eq:ordnryfld_act_wJ0}
S=\int d^4x \Big\{&\int d^4\th \tr\Big(\Ph\bar{\Ph} e^V + \bar{\Psi}\Psi e^{-V} -c^\prime V \Big) \nn \\
&+\Big[ \int d^2\th \tr\Big(\Xi\Ph\Psi +\Ph_0 \Ph J \Ph^T+\Psi_0\Psi^T J \Psi\Big)+\mathrm{(c.t.)}\Big]\Big\} ,\nn\\
&(c^\prime \in \mathbf{R}_{\neq 0}). 
\end{align}
$\Ph$ is an $N\times 2N$ chiral matrix field and $\Psi$ is a $2N \times N$ chiral matrix field. $\Ph_0$ and $\Psi_0$ are $N \times N$ chiral matrix fields, which are introduced as Lagrange multipliers. The Lagrangian of the action (\ref{eq:ordnryfld_act_wJ0}) is constrained by 
\begin{eqnarray}
&\Ph \Psi=0, \quad \mbox{(c.t)}=0, \label{eq:phpsi} \\
&\Ph J \Ph^T=0, \quad \mbox{(c.t)}=0, \label{eq:pjp} \\
&\Psi^T J \Psi=0, \quad \mbox{(c.t)}=0, \label{eq:hktpjtp1}
\end{eqnarray}
with the invariant tensor $J$ of $O(2N)$\footnote{We should remove the half of the result, which is related to the other half of the result by the parity since $J$ for $SO(2N)/U(N)$ allows the parity transformation of $O(2N)$ \cite{Lee:2017kaj,Arai:2011gg}.} or $USp(2N)$:
\begin{align} \label{eq:inv_ten_j}
J=\lt(
\begin{array}{cc}
0        &  I_N  \\
\ep I_N  &   0
\end{array}
\rt), \quad
\ep=\lt\{
\begin{array}{c}
1, \quad \mbox{for}~ SO(2N)/U(N) \\
-1, \quad \mbox{for}~ Sp(N)/U(N).
\end{array}
\rt. 
\end{align}
As $\Ph J \Ph^T$ and $\Psi^T J \Psi$ are symmetric (antisymmetric) for $SO(2N)$ ($USp(2N)$), $\Ph_0$ and $\Psi_0$ are symmetric (antisymmetric) rank-2 tensors:
\begin{align}
\Ph_0^T=\ep \Ph_0,\quad \Psi_0^T=\ep\Psi_0, \quad
\ep=\lt\{
\begin{array}{c}
1, \quad \mbox{for}~ SO(2N)/U(N) \\
-1, \quad \mbox{for}~ Sp(N)/U(N).
\end{array}
\rt. 
\end{align}
The $U(1)$ charge of $\Ph_0$ is $-2$ whereas the $U(1)$ charge of $\Psi_0$ is $+2$ to cancel out the $\Phi$ charge and the $\Psi$ charge respectively.

The chiral field that parametrises the base manifold is determined by the sign of the FI parameter $c^\prime$. In either $(c^\prime >0,\Psi=0=\bar{\Psi})$ case or $(c^\prime <0,\Ph=0=\bar{\Ph})$ case, the ${\mathcal{N}=2}$ nonlinear sigma models on $SO(2N)/U(N)$ and $Sp(N)/U(N)$ in the ${\mathcal{N}=1}$ superspace formalism (\ref{eq:ordnryfld_act_wJ0}) get reduced to the ${\mathcal{N}=1}$ nonlinear sigma models on $SO(2N)/U(N)$ and $Sp(N)/U(N)$ that are constructed in \cite{Higashijima:1999ki}. We choose the case $c^\pr =c > 0$ so that $\Phi$ parametrises only the base manifold. 

It should be noted that field $g$ in (\ref{eq:b=0}) and (\ref{eq:fg_constraints}), which parametrises the cotangent space as the fiber, is not constrained by the invariant tensor $J$ in (\ref{eq:inv_ten_j})\footnote{In \cite{Arai:2006gg}, the arctic superfield $\Upsilon(\z)=\varphi+\S \z +{\mathcal{O}(\z^2)}$ is constrained by $\Upsilon^T+\ep\Upsilon=0$. Therefore, the one-form field $\chi$, which is obtained by dualising $\S$, is also constrained by $\chi^T+\ep \chi=0$. }. As $g$ is not constrained by $J$, the bosonic component field of $\Psi$ should not contribute to the continuous vacuum. Therefore, the bosonic component field of the homogeneous $\Psi$ before the gauge fixing, also does not contribute to the vacuum.

The action of the mass-deformed ${\mathcal{N}}=2$ nonlinear sigma models on $SO(2N)/U(N)$ and $Sp(N)/U(N)$ can be obtained by introducing mass terms as it is done in (\ref{ea:n=2nlsmord}):
\begin{align} \label{eq:sosp_mass}
S= \int d^4x \Big\{&\int d^4\th \tr\Big(\Ph\bar{\Ph} e^V + \bar{\Psi}\Psi e^{-V}-c V \Big) \nn \\
&+\Big[ \int d^2\th \tr\Big(\Xi\Ph\Psi+\Ph M\Psi +\Ph_0\Ph J\Ph^T +\Psi_0\Psi^TJ\Psi\Big) +(\mbox{c.t.})\Big] \Big\}, \nn\\
&(c\in \mathbf{R}_{> 0}).
\end{align}
By introducing the mass terms, the most part of the continuous vacuum is lifted and the discrete vacua are left on the surface that is defined by the F-term constraints. As discussed before, the bosonic component of $\Psi$ should vanish at any vacuum. Therefore, the component field does not contribute to the BPS solutions which interpolate the discrete vacua. This observation is consistent with the results of \cite{Isozumi:2004jc,Isozumi:2004va,Eto:2006pg,Eto:2005cp,Eto:2005fm,Eto:2011cv}.
%

%%%%%%%%%%%%%%%%%%%%%%%%%%%%%%%%%%%%%%%%%%%%%%%%%%%%%%%%%%%%%%%%%%%%%%%%%%%%
\section{BPS solutions} \label{sec:bps_sol}
\setcounter{equation}{0}
%%%%%%%%%%%%%%%%%%%%%%%%%%%%%%%%%%%%%%%%%%%%%%%%%%%%%%%%%%%%%%%%%%%%%%%%%%%

In this section, we derive the mass-deformed nonlinear sigma models with complex masses on $SO(2N)/U(N)$ and $Sp(N)/U(N)$, which describe vacua, walls and three-pronged junctions and apply the moduli matrix formalism \cite{Isozumi:2004jc,Isozumi:2004va,Eto:2006pg,Eto:2005cp,Eto:2005fm,Shin:2018chr} to study the BPS objects. 

We are interested in the bosonic part of the action (\ref{eq:sosp_mass}). The superfields can be expanded by bosonic component fields as follows\footnote{We introduce the minus sign in $\Xi$ for ${\mcs}$ so that the vacuum labels in Section \ref{sec:so2n} and in Section \ref{sec:spn} are consistent with the labels of \cite{Lee:2017kaj,Arai:2018tkf}. }:
\begin{align}\label{ea:comp2}
&\Ph_{\sss{A}}^{~\sss{I}}(y)={\mca}_{\sss{A}}^{~\sss{I}}(y)+\th\th F_{\sss{A}}^{~\sss{I}}(y), ~ (y^\m=x^\m+i\th\s^\m\bar{\th}),  \nn\\
&\Psi_{\sss{I}}^{~\sss{A}}(y)={\mcb}_{\sss{I}}^{~\sss{A}}(y)+\th\th G_{\sss{I}}^{~\sss{A}}(y) , \nn\\
&V_{\sss{A}}^{~\sss{B}}(x)=2\th\s^\m\bar{\th}A_{\m{\sss{A}}}^{~~\sss{B}}(x)+\th\th\bar{\th}\bar{\th}D_{\sss{A}}^{~\sss{B}}(x), \nn\\
&\Xi_{\sss{A}}^{~\sss{B}}(y)=-{\mcs}_{\sss{A}}^{~\sss{B}}(y)+\th\th K_{\sss{A}}^{~\sss{B}}(y), \nn\\
&\Ph_0^{~\sss{AB}}(y)={\mca}_0^{~\sss{AB}}(y)+\th\th F_0^{~\sss{AB}}(y), \nn\\
&\Psi_{0\sss{AB}}(y)={\mcb}_{0\sss{AB}}(y)+\th\th G_{0\sss{AB}}(y), \nn\\
&(A=1,\cdots,N; I=1,\cdots,2N).
\end{align}
The bosonic part of the action  (\ref{eq:sosp_mass}) is
\begin{align} \label{eq:n=1_component_act2}
S=\int & d^4x \tr\Big(D_\m\mca\overline{D^\m\mca}
+\overline{D_\m\mcb} D^\m\mcb \nn\\
&-|\mca M -\mcs\mca+2\mcb_0\mcb^T J|^2 - |M\mcb-\mcb \mcs +2J\mca^T\mca_0|^2
\Big), \quad (\m=0,1,2,3),
\end{align}
with constraints
\begin{align}
&\mca\bar{\mca}-\bar{\mcb}\mcb-cI_M=0, \nn\\
&\mca\mcb=0, \quad \mbox{(c.t)=0},  \nn\\
&\mca J \mca^T=0, \quad \mbox{(c.t)=0}, \nn\\
&\mcb^T J \mcb=0, \quad \mbox{(c.t)=0}. 
\end{align}
The covariant derivatives are defined by
\begin{align}\label{eq:n=1_covariant_der}
D_\m\mca=\p_\m\mca-iA_\m \mca, \quad D_\m\mcb=\p_\m\mcb+i \mcb A_\m.
\end{align}

We set $\mcb=0=\bar{\mcb}$ to obtain the Lagrangian that describes vacua, walls and three-pronged junctions, since the fields do not contribute to the BPS solutions as discussed in Section \ref{sec:model} and in \cite{Isozumi:2004jc,Isozumi:2004va,Eto:2006pg,Eto:2005cp,Eto:2005fm,Eto:2011cv}. Then the Lagrangian that describes vacua, walls and junctions of the mass-deformed nonlinear sigma models on $SO(2N)/U(N)$ and $Sp(N)/U(N)$ in four dimensions is
\begin{align} \label{eq:n=1_component_lag_sosp}
{\mathcal{L}} =\tr\Big(D_\m\mca \overline{D^\m\mca} -|\mca M -\mcs\mca|^2 - |2J\mca^T\mca_0|^2 \Big),
\end{align}
with constraints
\begin{align} \label{eq:const_a}
&\mca\bar{\mca}-cI_M=0, \nn\\
&\mca J \mca^T=0, \quad \mbox{(c.t.)=0}.
\end{align}
The complex mass matrix $M$ and the complex matrix field $\mcs$ are diagonal by construction. 

The Lagrangian (\ref{eq:n=1_component_lag_sosp}) can be obtained by replacing the real-valued mass matrix and the scalar matrix field of the K\"{a}hler nonlinear sigma models on $SO(2N)/U(N)$ and $Sp(N)/U(N)$ \cite{Lee:2017kaj,Arai:2018tkf,Arai:2011gg} with complex-valued ones. This type of extension is applied to construct dyonic configurations with nonproportional charge vectors \cite{Eto:2011cv}.

The complex mass matrix is defined by a linear combination of the Cartan generators. The Cartan generators of $SO(2N)$ and $USp(2N)$ are
\begin{align} \label{eq:cartan_gen}
H_I=e_{{\sss{I}},{\sss{I}}}-e_{{\sss{N+I}},{\sss{N+I}}}, \quad (I=1,\cdots,N),
\end{align}
where $e_{I,I}~(e_{N+I,N+I})$ is a $2N\times 2N$ matrix of which the $(I,I)~((N+I,N+I))$ component is one.  By introducing vectors
\begin{align}
&\vec{l}:=(m_1+in_1,m_2+in_2,\cdots,m_{\sss{N}}+in_{\sss{N}}), \nn\\
&\vec{H}:=(H_1,H_2,\cdots,H_N),
\end{align}
with real-valued $m_i$ and $n_i$, $(i=1,\cdots,N)$, the mass matrix is formulated as
\begin{align}
M=\vec{l}\cdot\vec{H}.
\end{align}
$\mcs$ can be parametrised as
\begin{align}
\mcs=\mathrm{diag}(\s_1+i\ta_1,\s_2+i\ta_2,\cdots,\s_{\sss{N}}+i\ta_{\sss{N}}),
\end{align}
with real-valued $\s_i$ and $\ta_i$, $(i=1,\cdots,N)$.

The vacuum conditions of the Lagrangian (\ref{eq:n=1_component_lag_sosp}) are
\begin{align}
&\mca M -\mcs\mca=0,\quad \mbox{(c.t.)}=0, \nn\\
&\mca^T\mca_0=0, \quad \mbox{(c.t.)}=0.
\end{align}
Therefore, the vacuum solutions are labelled by
\begin{align}
&\lt(\s_1+i\ta_1,\s_2+i\ta_2,\cdots,\s_{\sss{N}}+i\ta_{\sss{N}}\rt)\nn\\
&\quad =\lt(\pm(m_1+in_1),\pm(m_2+in_2),\cdots,\pm(m_{\sss{N}}+in_{\sss{N}})\rt).
\end{align}
There are $2^{N-1}$ vacua in the mass-deformed nonlinear sigma model on $SO(2N)/U(N)$ as $J$ with $\ep=1$ in (\ref{eq:inv_ten_j}) is the $O(N)$ invariant tensor, and $2^N$ vacua in the mass-deformed nonlinear sigma model on $Sp(N)/U(N)$ \cite{Arai:2011gg}\footnote{The numbers of vacua of the mass-deformed nonlinear sigma models on the Hermitian symmetric spaces are the Euler characteristics of the spaces \cite{Eto:2011cv,Witten:1982df}.}.

The Lagrangian (\ref{eq:n=1_component_lag_sosp}) can be rewritten as
\begin{align} \label{eq:n=1_component_lag_sosp_real_ms}
{\mathcal{L}}=\tr\Big(D_\m\mca \overline{D^\m\mca}-\sum_{a=1,2}|\mca M_a -\S_a\mca|^2 - |2J\mca^T\mca_0|^2\Big),
\end{align}
by introducing real-valued matrices $M_a$ and $\S_a$, $(a=1,2)$
\begin{align}
&M=M_1+iM_2,  \nn\\
&\mcs=\S_1+i\S_2.
\end{align}  

We study junctions of the mass-deformed nonlinear sigma models on $SO(2N)/U(N)$ and $Sp(N)/U(N)$ with the Lagrangian (\ref{eq:n=1_component_lag_sosp_real_ms}). We are interested in static configurations, which are independent of the $x^3$-coordinate. We also assume that there is the Poincar\'{e} invariance on the worldvolume. So we fix $\p_0=\p_3=0$ and $A_0=A_3=0$. The energy density is
\begin{align} \label{eq:e_den}
{\mathcal{E}}=\tr\lt(\sum_{\a=1,2}\Big|D_\a\mca\mp\lt(\mca M_\a -\S_\a \mca \rt)\Big|^2
+ |2J\mca^T\mca_0|^2\rt) \pm {\mathcal{T}} \geq \pm{\mathcal{T}}, 
\end{align}
where the tension density is
\begin{align} \label{eq:ten_den}
{\mathcal{T}}=\tr\lt(\sum_{\a=1,2}\p_\a\lt(\mca M_\a \bar{\mca} \rt)\rt).
\end{align}
The index $\a=1,2$ is used for both codimensions and adjoint scalars as it is done in \cite{Eto:2005cp}. The energy density (\ref{eq:e_den}) and the tension density (\ref{eq:ten_den}) are constrained by (\ref{eq:const_a}).

The (anti)BPS equation is
\begin{align}
D_\a\mca\mp\lt(\mca M_\a -\S_\a \mca \rt)=0,\quad (\a=1,2).
\end{align}
We choose the upper sign for the BPS equation. We apply the moduli matrix formalism \cite{Isozumi:2004jc,Isozumi:2004va,Eto:2006pg,Eto:2005cp,Eto:2005fm} to solve the BPS equation. 
The BPS solution is
\begin{align} \label{eq:bps_sol_a}
\mca=S^{-1}H_0e^{M_1 x^1 +M_2 x^2},
\end{align}
with a relation
\begin{align} \label{eq:sps}
S^{-1}\p_\a S:=\S_\a-iA_\a,\quad (\a=1,2).
\end{align}
The coefficient matrix $H_0$ is the moduli matrix. The constraints in (\ref{eq:const_a}) are
\begin{align}
&S\bar{S}=\frac{1}{c}H_0 e^{2M_1 x^1 +2M_2 x^2} \bar{H}_0, \label{eq:const_h1}\\
&H_0 J H_0^T =0,\quad \mbox{(c.t)}=0. \label{eq:const_h2}
\end{align}
The BPS solution (\ref{eq:bps_sol_a}), $\S_\a$ and $A_\a$ in (\ref{eq:sps}) are invariant under the transformation:
\begin{align} \label{eq:equivcls}
H_0^\prime=V H_0, \quad S^\prime=V S, \quad V\in GL(N,\mathbf{C}).
\end{align}
The equivalence class of $(S,H_0)$ is the worldvolume symmetry of the moduli matrix formalism. The first equation of (\ref{eq:equivcls}) and (\ref{eq:const_h2}) show that the moduli matrices $H_0$'s parametrise $SO(2N)/U(N)$ and $Sp(N)/U(N)$ respectively \cite{Arai:2011gg}.

Walls are built by elementary walls, which are constructed by the simple root generators of the global symmetry. Therefore, the elementary walls are identified with the simple roots \cite{Lee:2017kaj,Arai:2018tkf,Sakai:2005sp}. We summarise the simple root generators $E_i$, $(i=1,\cdots,N)$ and the simple root $\vec{\a}_i$ of $SO(2N)$ and $USp(2N)$ \cite{Eto:2011cv,Isaev:2018xcg}: \\

\bul $SO(2N)$
\begin{align} \label{eq:so2n_rts}
&E_i=e_{i,i+1}-e_{i+{\sss{N}}+1,i+{\sss{N}}},\quad (i=1,\cdots,N-1), \nn\\
&E_N=e_{{\sss{N}}-1,2{\sss{N}}}-e_{{\sss{N}},2{\sss{N}}-1}, \nn\\
&\vec{\a}_i=\hat{e}_i-\hat{e}_{i+1}, \nn\\
&\vec{\a}_{\sss{N}}=\hat{e}_{{\sss{N}}-1}+\hat{e}_{\sss{N}}.
\end{align}

\bul $USp(2N)$ 
\begin{align}\label{eq:uspn_rts}
&E_i=e_{i,i+1}-e_{i+{\sss{N}}+1,i+{\sss{N}}}, \quad (i=1,\cdots,N-1), \nn\\
&E_N=e_{{\sss{N}},2{\sss{N}}}, \nn\\
&\vec{\a}_i=\hat{e}_i-\hat{e}_{i+1}, \nn\\
&\vec{\a}_{\sss{N}}=2\hat{e}_{\sss{N}}.
\end{align}

%%%%%%%%%%%%%%%%%%%%%%%%%%%%%%%%%%%%%%%%%%%%%%%%%%%%%%%%%%%%%%%%%%%%%%%%%%%%
\section{Three-pronged junctions of the mass-deformed nonlinear sigma models on $SO(8)/U(4)$ and $Sp(3)/U(3)$ } \label{sec:3pronged_junct}
\setcounter{equation}{0}
%%%%%%%%%%%%%%%%%%%%%%%%%%%%%%%%%%%%%%%%%%%%%%%%%%%%%%%%%%%%%%%%%%%%%%%%%%%

The moduli matrix formalism is applied to three-pronged junctions of the mass-deformed nonlinear sigma model on the complex projective space, which is an Abelian gauge theory \cite{Eto:2005cp}, and to three-pronged junctions of the mass-deformed nonlinear sigma model on the Grassmann manifold $G_{N_F,N_C}$, which is a non-Abelian gauge theory for $N_C\geq 2$ \cite{Eto:2005fm}. In the moduli matrix formalism for the complex projective space and the Grassmann manifold, vacua are labelled by assigning a nonvanishing flavour number for each colour component \cite{Isozumi:2004jc,Isozumi:2004va,Eto:2006pg}. There are two types of three-pronged junctions. An Abelian three-pronged junction divides three vacua that differ by one label number: $\lra{\cdots A}$,$\lra{\cdots B}$,$\lra{\cdots C}$. A non-Abelian three-pronged junction divides three vacua that differ by two label numbers: $\lra{\cdots AB}$,$\lra{\cdots BC}$,$\lra{\cdots AC}$. Abelian three-pronged junctions exist both in Abelian gauge theories and in non-Abelian gauge theories whereas non-Abelian three-pronged junctions exist only in non-Abelian gauge theories \cite{Eto:2005cp,Eto:2005fm}.

In \cite{Eto:2005cp}, the moduli matrix formalism is directly applied to three-pronged junctions of the mass-deformed nonlinear sigma models on the complex projective spaces since vacua have only one colour component so that neighbouring vacua can be easily identified. Two-dimensional diagrams or polyhedron diagrams are proposed to describe vacua and BPS objects of the mass-deformed nonlinear sigma models on the complex projective space. Vertices, edges and triangular faces of the polyhedron diagrams correspond to vacua, walls and three-pronged junctions.

The moduli matrix formalism for three-pronged junctions accompanies technical difficulties in the mass-deformed nonlinear sigma model on the Grassmann manifold. Neighbouring vacua that are interpolated by walls should be identified and $S^{-1}$ of (\ref{eq:bps_sol_a}) should be obtained from the matrix $S\bar{S}$ in (\ref{eq:const_h1}), which is not in general diagonal for non-Abelian junctions. In \cite{Eto:2005fm}, the Grassmann manifold $G_{N_F,N_C}$ is embedded in the complex projective space $\mathbf{C}P^{_{\sss{N_F}} C_{\sss{N_C}}-1}$ by the Pl\"{u}cker embedding resolving the difficulties. This method is efficient, however, its disadvantage is that it cannot be directly applied to the quadrics in the Grassmann manifold since it is not yet known how to impose the quadratic constraints to the Pl\"{u}cker embedding.

It is shown in \cite{Shin:2018chr} that we can apply the pictorial representation, which is proposed in \cite{Lee:2017kaj}, to the mass-deformed nonlinear sigma model on the Grassmann manifold and reformulate the diagrams for vacua and elementary walls in the pictorial representation to build polyhedron diagrams, which are similar to the polyhedron diagrams that are proposed in \cite{Eto:2005cp} to describe vacua, walls and three-pronged junctions of the mass-deformed nonlinear sigma models on the complex projective space, which are Abelian gauge theories. Positions of non-Abelian three-pronged junctions as well as Abelian three-pronged junctions are calculated from the diagrams.

We show in this paper for the first time that we can apply the moduli matrix formalism to three-pronged junctions of the mass-deformed nonlinear sigma models on $SO(2N)/U(N)$ and $Sp(N)/U(N)$ that are constructed in the ${\mathcal{N}}=1$ superspace formalism. The models are non-Abelian gauge theories for $N\geq 4$ and $N\geq 3$ respectively. We construct three-pronged junctions on $SO(8)/U(4)$ and $Sp(3)/U(3)$ by making use of the diagram method in the pictorial representation that is proposed in \cite{Shin:2018chr}. As a sequel to this paper, the approach that is proposed in this paper is applied to the mass-deformed nonlinear sigma models on $SO(2N)/U(N)$ and $Sp(N)/U(N)$ with generic $N$ \cite{Kim:2020obf}.

%%%%%%%%%%%%%%%%%%%%%%%%%%%%%%%%%%%%%%%%%%%%%%%%%%%%%%%%%%%%%%%%%%%%%%%%%%%%
\subsection{Three-pronged junctions of the mass-deformed nonlinear sigma model on $SO(8)/U(4)$} \label{sec:so2n}
%\setcounter{equation}{0}
%%%%%%%%%%%%%%%%%%%%%%%%%%%%%%%%%%%%%%%%%%%%%%%%%%%%%%%%%%%%%%%%%%%%%%%%%%%

In this subsection,  we review the pictorial representation that is proposed in \cite{Lee:2017kaj}, and apply the diagram method in the pictorial representation that is proposed in \cite{Shin:2018chr} to construct three-pronged junctions of the mass-deformed nonlinear sigma models on $SO(2N)/U(N)$. To study vacua and walls, we can reduce the model, which is discussed in Section \ref{sec:bps_sol} by setting $M_2=0$ and $\S_2=0$. As we are interested in generic mass parameters, we can set $m_i>m_{i+1}$ without loss of generality. We label the vacua in descending order \cite{Lee:2017kaj}:
\begin{align} \label{eq:so_vac_rule}
&(\s_1,\s_2,\cdots,\s_{N-1},\s_N)=(m_1,m_2,\cdots,m_{N-1},m_N),  \nn\\
&(\s_1,\s_2,\cdots,\s_{N-1},\s_N)=(m_1,m_2,\cdots,-m_{N-1},-m_N),  \nn\\
&\quad \vdots \nn\\
&(\s_1,\s_2,\cdots,\s_{N-1},\s_N)=(\pm m_1,-m_2,\cdots,-m_{N-1},-m_N),
\end{align}
where the sign $\pm$ is $+$ for odd $N$ and $-$ for even $N$.

Let $\lra{\cdot}$ denote a vacuum, and $\lra{\cdot \leftarrow \cdot}$ or $\lra{\cdot \leftrightarrow \cdot}$ denote a wall. Elementary wall $\lra{A\leftarrow B}$ that interpolates vacuum $\lra{A}$ and vacuum $\lra{B}$ of the mass-deformed nonlinear sigma model on $SO(2N)/U(N)$ is defined by the following relation:
\begin{align}\label{eq:ele_alg}
2c[M,E_i]=2c(\vec{m}\cdot \vec{\a}_i)E_i=T_{\lra{A\leftarrow B}} E_i, \quad (i=1,\cdots,N).
\end{align}
$E_i$ is the positive root generator of the simple root of $SO(2N)$, which corresponds to the elementary wall operator of the moduli matrix formalism. Constant $c$ is the FI parameter of the Lagrangian (\ref{eq:sosp_mass}). Elementary wall $H_{0\lra{A\leftarrow B}}=H_{0\lra{A}}e^{E_i(r)}$ with $E_i(r)\equiv e^r E_i$, $(r\in\mathbf{C})$ is labelled by simple root $\vec{\a}_i$ in the pictorial representation. The elementary wall has tension $T_{\lra{A\leftarrow B}}$. Let $\vec{g}_{\lra{A\leftarrow B}}$ denote the elementary wall, which satisfies the relation (\ref{eq:ele_alg}):
\begin{align}
\vec{g}_{\lra{A\leftarrow B}} \equiv 2c\vec{\a}_i, \quad (i=1,\cdots,N).
\end{align}
A compressed wall of level $l$ is  
\begin{align}
\vec{g}_{\lra{\cdots}} = 2c\vec{\a}_{i_1} + 2c\vec{\a}_{i_2} + \cdots + 2c\vec{\a}_{i_{l+1}},\quad
(i_m=1,\cdots,N;~m\leq l+1).
\end{align}
A pair of penetrable walls are orthogonal:
\begin{align}
\vec{g}_{\lra{\cdots}}\cdot \vec{g}_{\lra{\cdots}}=0.
\end{align}
Elementary walls, compressed walls and multiwalls of the mass-deformed nonlinear sigma models on $SO(2N)/U(N)$ are discussed in \cite{Lee:2017kaj,Arai:2011gg}. 

\begin{figure}[ht!]
\begin{center}
\includegraphics[width=7cm,clip]{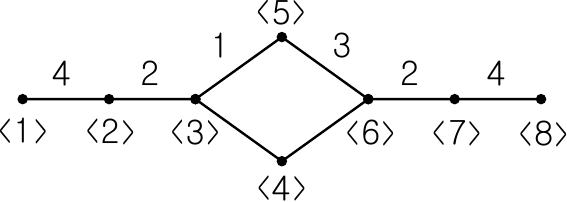}
\end{center}
 \caption{ Vacua and elementary walls of the mass-deformed nonlinear sigma model on $SO(8)/U(4)$ \cite{Lee:2017kaj}. The numbers in brackets indicate the vacuum labels. The numbers without brackets indicate the subscript $i$'s of simple roots $\vec{\a}_i$, $(i=1,\cdots,4)$.}
 \label{fig:so8_wall}
\end{figure}

We present the diagram for the vacua and the elementary walls of the mass-deformed nonlinear sigma model on $SO(8)/U(4)$ \cite{Lee:2017kaj} in Figure \ref{fig:so8_wall}. Vertices and segments correspond to vacua and elementary walls. The parallelogram presents two pairs of penetrable walls. The facing sides of the parallelogram are the same vectors whereas the adjacent sides of the parallelogram are orthogonal vectors.

Let us look at sector $\lt\{\lra{1},\lra{2},\lra{3} \rt\}$. The vacua are labelled by the rule (\ref{eq:so_vac_rule}):
\begin{align}
&\lra{1}~:~(\s_1,\s_2,\s_3,\s_4)=(m_1,m_2,m_3,m_4), \nn\\
&\lra{2}~:~(\s_1,\s_2,\s_3,\s_4)=(m_1,m_2,-m_3,-m_4), \nn\\
&\lra{3}~:~(\s_1,\s_2,\s_3,\s_4)=(m_1,-m_2,m_3,-m_4).
\end{align}
We can learn from Figure \ref{fig:so8_wall} that the elementary wall operators of $\lra{1\leftarrow 2}$ and $\lra{2\leftarrow 3}$ are $E_4$ and $E_2$ of (\ref{eq:so2n_rts}) respectively. We can also learn from (\ref{eq:so2n_rts}) that $[E_4,E_2]\neq0$. Therefore, there exists compressed wall $\lra{1\leftarrow 3}$. The moduli matrices of the single walls that interpolate vacua $\{\lra{1},\lra{2},\lra{3}\}$ are
\begin{align} \label{eq:so8_single_walls}
&H_{0\lra{1\leftarrow 2}}=H_{0\lra{1}}e^{e^rE_4}, \nn\\
&H_{0\lra{2\leftarrow 3}}=H_{0\lra{2}}e^{e^rE_2}, \nn\\
&H_{0\lra{1\leftarrow 3}}=H_{0\lra{1}}e^{e^r\lt[E_4,E_2\rt]}.
\end{align}

\begin{figure}[ht!]
\begin{center}
\includegraphics[width=7cm,clip]{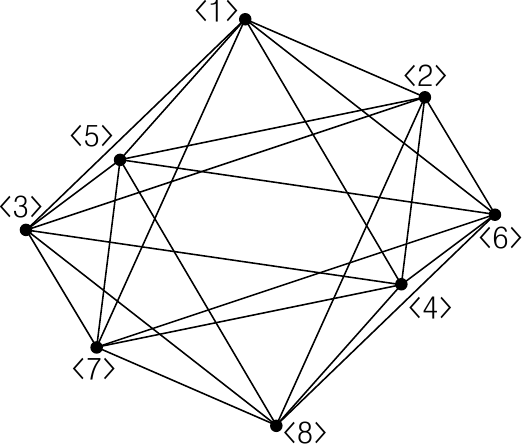}
\end{center}
 \caption{ Vacua, walls and three-pronged junctions of the mass-deformed nonlinear sigma models on $SO(8)/U(4)$. Vertices, edges and triangular faces correspond to vacua, walls and three-pronged junctions.}
 \label{fig:so8_junctions}
\end{figure}

We study three-pronged junctions. We turn on mass matrix $M_2$ and matrix field $\S_2$ and set $m_1+in_1\neq m_2+in_2\neq \cdots \neq m_{\sss{N}}+in_{\sss{N}}$. From (\ref{eq:so2n_rts}), we find that
\begin{align}
&\vec{\a}_1\cdot(\vec{\a}_2+\vec{\a}_4)\neq 0, \label{eq:so_comp_124}\\
&\vec{\a}_3\cdot(\vec{\a}_2+\vec{\a}_4)\neq 0. \label{eq:so_comp_324}
\end{align}
Eq.(\ref{eq:so_comp_124}) shows that there exist compressed walls $\lra{1 \leftarrow 5}$ and $\lra{4 \leftarrow 8}$. Eq.(\ref{eq:so_comp_324}) shows that there exist compressed walls $\lra{1\leftarrow 4}$ and $\lra{5 \leftarrow 8}$. We reformulate the diagram in Figure \ref{fig:so8_wall} by connecting vacua to produce the diagram presented in Figure \ref{fig:so8_junctions}. Vertices, edges and triangular faces correspond to vacua, walls and three-pronged junctions. 

We study the three-pronged junction that divides vacua $\{\lra{1},\lra{2},\lra{3} \}$. The vacua are labelled by the rule (\ref{eq:so_vac_rule}):
\begin{align} \label{eq:sovac123}
&\lra{1} ~:~(\s_1+i\ta_1,\s_2+i\ta_2,\s_3+i\ta_3,\s_4+i\ta_4) \nn\\
&\quad\quad\quad =(m_1+in_1,m_2+in_2,m_3+in_3,m_4+in_4),  \nn\\
&\lra{2} ~:~(\s_1+i\ta_1,\s_2+i\ta_2,\s_3+i\ta_3,\s_4+i\ta_4) \nn\\
&\quad\quad\quad =(m_1+in_1,m_2+in_2,-m_3-in_3,-m_4-in_4), \nn\\
&\lra{3} ~:~(\s_1+i\ta_1,\s_2+i\ta_2,\s_3+i\ta_3,\s_4+i\ta_4) \nn\\
&\quad\quad\quad =(m_1+in_1,-m_2-in_2,m_3+in_3,-m_4-in_4).
\end{align}
We apply the worldvolume symmetry transformation (\ref{eq:equivcls}) to the moduli matrices of walls (\ref{eq:so8_single_walls}) to produce a general description ($h_{\sss{AI}}=\exp(a_{\sss{AI}}+ib_{\sss{AI}})$; $a_{\sss{AI}},b_{\sss{AI}}\in\mathbf{R}$):

\begin{align}
&H_{0\lra{1\leftrightarrow 2}}=
\lt(
\begin{array}{cccccccc}
1    &   ~   &  ~   &   ~        &    0   &   ~   &    ~    &    ~   \\
~    &   1   &  ~   &   ~        &    ~   &   0   &    ~    &    ~   \\
~    &   ~   &  h_{33}   &   ~   &    ~   &   ~   &    0    &    h_{38}   \\
~    &   ~   &  ~   &   h_{33}   &    ~   &   ~   &    -h_{38}    &    0   
\end{array}
\rt), \nn\\
&\quad \nn\\
&H_{0\lra{2\leftrightarrow 3}}=
\lt(
\begin{array}{cccccccc}
1    &   ~        &  ~       &   ~   &    0   &   ~      &    ~    &    ~   \\
~    &   h_{22}   &  h_{23}  &   ~   &    ~   &   0      &    ~    &    ~   \\
~    &   ~        &  0       &   ~   &    ~   &   -h_{23}          &    h_{22}    &    ~   \\
~    &   ~        &  ~       &   0   &    ~   &   ~      &   ~     &    1   
\end{array}
\rt), \nn\\
&\quad \nn\\
&H_{0\lra{3\leftrightarrow 1}}=
\lt(
\begin{array}{cccccccc}
1    &   ~   &  ~   &   ~   &    0   &   ~   &    ~    &    ~   \\
~    &   h_{22}   &  ~   &   ~   &    ~   &   0   &    ~    &   h_{28}   \\
~    &   ~   &  1   &   ~   &    ~   &   ~   &    0    &    ~   \\
~    &   ~   &  ~   &   h_{22}   &    ~   &  -h_{28}   &   ~    &    0   
\end{array}
\rt).
\end{align}
As the moduli matrices have the worldvolume symmetry (\ref{eq:equivcls}), only one of $h_{\sss{AI}}$ parameters or the ratio of the two parameters in each moduli matrix is independent.

The wall solutions that interpolate $\lt\{\lra{1},\lra{2}\rt\}$, $\lt\{\lra{2},\lra{3}\rt\}$ and $\lt\{\lra{3},\lra{1}\rt\}$ are obtained by (\ref{eq:bps_sol_a}) with (\ref{eq:const_h1}) as follows:
\\~\\
$\noindent\bullet$$\Big\{\lra{1},\lra{2}\Big\}$
\begin{align} \label{eq:so8_12_wall}
%&\bullet\Big\{\lra{1},\lra{2}\Big\}\nn\\
&\mca_{\lra{1\leftrightarrow 2}}=
\lt(
\begin{array}{cccccccc}
\sqrt{c}    &   ~   &  ~   &   ~   &    0   &   ~   &    ~    &    ~   \\
~    &   \sqrt{c}   &  ~   &   ~   &    ~   &   0   &    ~    &    ~   \\
~    &   ~   &  \frac{p_{33}}{\sqrt{s_3}}   &   ~   &    ~   &   ~   &    0    &   \frac{p_{38}}{\sqrt{s_3}}  \\
~    &   ~   &  ~   &  \frac{p_{44}}{\sqrt{s_4}}   &    ~   &   ~  &   \frac{p_{47}}{\sqrt{s_4}}    &    0   
\end{array}
\rt), \nn\\
&p_{33}=\exp(m_3x^1+n_3x^2+a_{33}+ib_{33}), \nn\\
&p_{38}=\exp(-m_4x^1-n_4x^2+a_{38}+ib_{38}), \nn\\
&p_{44}=\exp(m_4x^1+n_4x^2+a_{33}+ib_{33}),   \nn\\
&p_{47}=\exp(-m_3x^1-n_3x^2+a_{38}+ib_{38}+i\pi), \nn\\
&s_3=\frac{1}{c}\lt[\exp(2m_3x^1+2n_3x^2+2a_{33})+\exp(-2m_4x^1-2n_4x^2+2a_{38})\rt], \nn\\
&s_4=\frac{1}{c}\lt[\exp(2m_4x^1+2n_4x^2+2a_{33})+\exp(-2m_3x^1-2n_3x^2+2a_{38})\rt].
\end{align}
\\~\\
$\noindent\bullet$$\Big\{\lra{2},\lra{3}\Big\}$
\begin{align}\label{eq:so8_23_wall}
&\mca_{\lra{2\leftrightarrow 3}}=
\lt(
\begin{array}{cccccccc}
\sqrt{c}    &   ~   &  ~   &   ~   &    0   &   ~   &    ~    &    ~   \\
~    &   \frac{q_{22}}{\sqrt{t_2}}   &  \frac{q_{23}}{\sqrt{t_2}}  &   ~   &    ~   &   0   &    ~    &    ~   \\
~    &   ~   &  0   &   ~   &    ~   &  \frac{q_{36}}{\sqrt{t_3}}   &    \frac{q_{37}}{\sqrt{t_3}}    &   ~ \\
~    &   ~   &  ~   &  0   &    ~   &  ~  &   ~    &    \sqrt{c}   
\end{array}
\rt), \nn\\
&q_{22}=\exp(m_2x^1+n_2x^2+a_{22}+ib_{22}), \nn\\
&q_{23}=\exp(m_3x^1+n_3x^2+a_{23}+ib_{23}), \nn\\
&q_{36}=\exp(-m_2x^1-n_2x^2+a_{23}+ib_{23}+i\pi),   \nn\\
&q_{37}=\exp(-m_3x^1-n_3x^2+a_{22}+ib_{22}), \nn\\
&t_2=\frac{1}{c}\lt[\exp(2m_2x^1+2n_2x^2+2a_{22})+\exp(2m_3x^1+2n_3x^2+2a_{23})\rt], \nn\\
&t_3=\frac{1}{c}\lt[\exp(-2m_2x^1-2n_2x^2+2a_{23})+\exp(-2m_3x^1-2n_3x^2+2a_{22})\rt].
\end{align}
\\~\\
$\noindent\bullet$$\Big\{\lra{3},\lra{1}\Big\}$
\begin{align} \label{eq:so8_13_wall}
&\mca_{\lra{3\leftrightarrow 1}}=
\lt(
\begin{array}{cccccccc}
\sqrt{c}    &   ~   &  ~   &   ~   &    0   &   ~   &    ~    &    ~   \\
~    &   \frac{r_{22}}{\sqrt{u_2}}   &  ~  &   ~   &    ~   &   0   &    ~    &   \frac{r_{28}}{\sqrt{u_2}}   \\
~    &   ~   &  \sqrt{c}   &   ~   &    ~   & ~   &    0   &   ~ \\
~    &   ~   &  ~   &  \frac{r_{44}}{\sqrt{u_4}}   &    ~   &  \frac{r_{46}}{\sqrt{u_4}} &   ~    &   0
\end{array}
\rt), \nn\\
&r_{22}=\exp(m_2x^1+n_2x^2+a_{22}+ib_{22}), \nn\\
&r_{28}=\exp(-m_4x^1-n_4x^2+a_{28}+ib_{28}), \nn\\
&r_{44}=\exp(m_4x^1+n_4x^2+a_{22}+ib_{22}),   \nn\\
&r_{46}=\exp(-m_2x^1-n_2x^2+a_{28}+ib_{28}+i\pi), \nn\\
&u_2=\frac{1}{c}\lt[\exp(2m_2x^1+2n_2x^2+2a_{22})+\exp(-2m_4x^1-2n_4x^2+2a_{28})\rt], \nn\\
&u_4=\frac{1}{c}\lt[\exp(2m_4x^1+2n_4x^2+2a_{22})+\exp(-2m_2x^1-2n_2x^2+2a_{28})\rt].
\end{align}
As mentioned previously, only one of $a_{\sss{AI}}$ parameters is independent in each wall solution.

The position of wall $\lra{1 \leftrightarrow 2}$ is $\mathrm{Re}(p_{33})=\mathrm{Re}(p_{38})$ and $\mathrm{Re}(p_{44})=\mathrm{Re}(p_{47})$:
\begin{align} \label{eq:so_wall_12_pos}
(m_3+m_4)x^1+(n_3+n_4)x^2+(a_{33}-a_{38})=0.
\end{align}
The position of wall $\lra{2 \leftrightarrow 3}$ is $\mathrm{Re}(q_{22})=\mathrm{Re}(q_{23})$ and $\mathrm{Re}(q_{36})=\mathrm{Re}(q_{37})$:
\begin{align} \label{eq:so_wall_23_pos}
(m_2-m_3)x^1+(n_2-n_3)x^2+(a_{22}-a_{23})=0.
\end{align}
The position of wall $\lra{3 \leftrightarrow 1}$ is $\mathrm{Re}(r_{22})=\mathrm{Re}(r_{28})$ and $\mathrm{Re}(r_{44})=\mathrm{Re}(r_{46})$:
\begin{align} \label{eq:so_wall_13_pos}
(m_2+m_4)x^1+(n_2+n_4)x^2+(a_{22}-a_{28})=0.
\end{align}
There is a consistency condition:
\begin{align}
a_{23}-a_{33}=a_{28}-a_{38}.
\end{align}
Therefore, there are two independent $a_{\sss{AI}}$ parameters to describe the junction position as expected.
We can compute the position of the junction from (\ref{eq:so_wall_12_pos}), (\ref{eq:so_wall_23_pos}) and (\ref{eq:so_wall_13_pos}). The position of the junction that divides the vacua $\{\lra{1},\lra{2},\lra{3}\}$ in (\ref{eq:sovac123}) is
\begin{align}
(x,y)=\lt(\frac{S_1}{S_3},\frac{S_2}{S_3}\rt),
\end{align}
\begin{align}
&S_1=(-a_{33}+a_{38})n_2+(a_{22}-a_{28})n_3 + (a_{22}-a_{23})n_4, \nn\\
&S_2=(a_{33}-a_{38})m_2+(-a_{22}+a_{28})m_3+(-a_{22}+a_{23})m_4, \nn\\
&S_3=-(m_2-m_3)(n_3+n_4)+(m_3+m_4)(n_2-n_3).
\end{align}
%

%%%%%%%%%%%%%%%%%%%%%%%%%%%%%%%%%%%%%%%%%%%%%%%%%%%%%%%%%%%%%%%%%%%%%%%%%%%%
\subsection{Three-pronged junctions of the mass-deformed nonlinear sigma model on $Sp(3)/U(3)$} \label{sec:spn}
%\setcounter{equation}{0}
%%%%%%%%%%%%%%%%%%%%%%%%%%%%%%%%%%%%%%%%%%%%%%%%%%%%%%%%%%%%%%%%%%%%%%%%%%%

In this subsection, we review the pictorial representation \cite{Arai:2018tkf} and apply the diagram method \cite{Shin:2018chr} to the mass-deformed nonlinear sigma models on $Sp(N)/U(N)$. To study vacua and  walls, we reduce the model by setting $M_2=0$ and $\S_2=0$. We also set $m_i>m_{i+1}$ without loss of generality. We label the vacua in descending order \cite{Arai:2018tkf}:
\begin{align}\label{eq:spn_vac_label}
&(\s_1,\s_2,\cdots,\s_{N-1},\s_N)=(m_1,m_2,\cdots,m_{N-1},m_N),  \nn\\
&(\s_1,\s_2,\cdots,\s_{N-1},\s_N)=(m_1,m_2,\cdots,m_{N-1},-m_N),  \nn\\
&(\s_1,\s_2,\cdots,\s_{N-1},\s_N)=(m_1,m_2,\cdots,-m_{N-1},m_N),  \nn\\
&(\s_1,\s_2,\cdots,\s_{N-1},\s_N)=(m_1,m_2,\cdots,-m_{N-1},-m_N),  \nn\\
&\quad \vdots \nn\\
&(\s_1,\s_2,\cdots,\s_{N-1},\s_N)=(m_1,-m_2,\cdots,-m_{N-1},-m_N), \nn\\
&(\s_1,\s_2,\cdots,\s_{N-1},\s_N)=(-m_1,m_2,\cdots,m_{N-1},m_N), \nn\\
&\quad \vdots \nn\\
&(\s_1,\s_2,\cdots,\s_{N-1},\s_N)=(-m_1,-m_2,\cdots,-m_{N-1},-m_N).
\end{align}

The elementary wall vectors should be defined by scaled simple roots since the lengths of the $USp(2N)$ simple roots are not all the same. Elementary wall $\lra{A\leftarrow B}$ that interpolates vacuum $\lra{A}$ and vacuum $\lra{B}$ of the mass-deformed nonlinear sigma models on $Sp(N)/U(N)$ is defined by the following relations:
\begin{align}
&2c[M,E_i]=2c(\vec{m}\cdot \vec{\a}_i)E_i = T_{\lra{A\leftarrow B}} E_i, \quad (i=1,\cdots,N-1), \nn\\
&c[M,E_N]=c(\vec{m}\cdot \vec{\a}_N)E_N = T_{\lra{A\leftarrow B}} E_N.
\end{align}
The corresponding elementary wall is
\begin{align}
&\vec{g}_{\lra{A\leftarrow B}} \equiv 2c\vec{\a}_i, \quad (i=1,\cdots,N-1), \nn\\
&\vec{g}_{\lra{A\leftarrow B}} \equiv c\vec{\a}_N.
\end{align}
A compressed wall of level $l$ is
\begin{align}
\vec{g}_{\lra{\cdots}}=2c\vec{\a}_{i_1}+2c\vec{\a}_{i_2}+\cdots +2c\vec{\a}_{i_{l+1}},\quad (i_m \leq N-1;~m\leq l+1),
\end{align}
for the simple root vectors of the same length. There is a compressed wall sector of unequal length simple roots:
\begin{align}
\vec{g}_{\lra{\cdots}}=2c\vec{\a}_{N-1}+c\vec{\a}_N.
\end{align}
This is the distinguishing feature of the mass-deformed nonlinear sigma models on $Sp(N)/U(N)$. A pair of penetrable walls are orthogonal.
Elementary walls, compressed walls and multiwalls of the mass-deformed nonlinear sigma models on $Sp(N)/U(N)$ are discussed in \cite{Arai:2018tkf}. 

\begin{figure}[ht!]
\begin{center}
\includegraphics[width=7cm,clip]{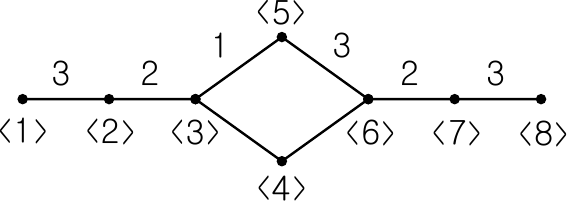}
\end{center}
 \caption{ Vacua and elementary walls of the mass-deformed nonlinear sigma model on $Sp(3)/U(3)$ \cite{Arai:2018tkf}. The numbers in brackets indicate the vacuum labels. The numbers without brackets indicate the subscript $i$'s of simple roots $\vec{\a}_i$, $(i=1,\cdots,3)$.}
 \label{fig:sp3_wall}
\end{figure}

We present the diagram for the vacua and the elementary walls of the mass-deformed nonlinear sigma model on $Sp(3)/U(3)$ \cite{Arai:2018tkf} in Figure \ref{fig:sp3_wall}. As before, vertices and segments correspond to vacua and elementary walls. The parallelogram presents two pairs of penetrable walls. The facing sides of the parallelogram are the same vectors whereas the adjacent sides of the parallelogram are orthogonal vectors.

Let us look at sector $\lt\{\lra{1},\lra{2},\lra{3} \rt\}$. The vacua are labelled by the rule (\ref{eq:spn_vac_label}):
\begin{align}
&\lra{1}~:~(\s_1,\s_2,\s_3)=(m_1,m_2,m_3), \nn\\
&\lra{2}~:~(\s_1,\s_2,\s_3)=(m_1,m_2,-m_3), \nn\\
&\lra{3}~:~(\s_1,\s_2,\s_3)=(m_1,-m_2,m_3).
\end{align}
We can learn from Figure \ref{fig:sp3_wall} that the elementary wall operators of $\lra{1\leftarrow 2}$ and $\lra{2\leftarrow 3}$ are $E_3$ and $E_2$ of (\ref{eq:uspn_rts}). We can also learn from (\ref{eq:uspn_rts}) that $[[E_3,E_2],E_2]\neq0$. Therefore, there exists compressed wall $\lra{1\leftarrow 3}$. The moduli matrices of the single walls that interpolate vacua $\{\lra{1},\lra{2},\lra{3}\}$ are
\begin{align} \label{eq:sp3_single_walls}
&H_{0\lra{1\leftarrow 2}}=H_{0\lra{1}}e^{e^rE_3}, \nn\\
&H_{0\lra{2\leftarrow 3}}=H_{0\lra{2}}e^{e^rE_2}, \nn\\
&H_{0\lra{1\leftarrow 3}}=H_{0\lra{1}}e^{e^r\lt[\lt[E_3,E_2\rt],E_2\rt]}.
\end{align}

\begin{figure}[ht!]
\begin{center}
\includegraphics[width=7cm,clip]{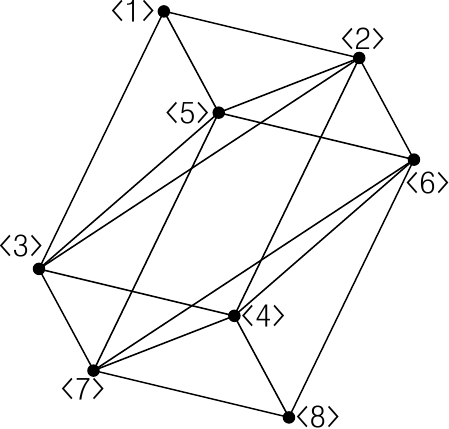}
\end{center}
 \caption{ Vacua, walls and three-pronged junctions of the mass-deformed nonlinear sigma models on $Sp(3)/U(3)$. Vertices, edges and triangular faces correspond to vacua, walls and three-pronged junctions.}
 \label{fig:sp3_junctions}
\end{figure}

We study three-pronged junctions. We turn on $M_2$ and $\S_2$, and set $m_1+in_1\neq m_2+in_2\neq \cdots \neq m_{\sss{N}}+in_{\sss{N}}$. From (\ref{eq:uspn_rts}), we find that 
\begin{align}
&(2\vec{\a}_2+\vec{\a}_3)\cdot\vec{\a}_3 = 0, \label{eq:sp_comp_233}\\
&(2\vec{\a}_1+2\vec{\a}_2+\vec{\a}_3)\cdot\vec{\a}_3 = 0. \label{eq:sp_comp_1233}
\end{align}
Eq.(\ref{eq:sp_comp_233}) shows that there do not exist compressed walls $\lra{1 \leftarrow 4}$ and $\la 5 \leftarrow 8 \ra$. Eq.(\ref{eq:sp_comp_1233}) shows that there do not exist compressed walls $\lra{1 \leftarrow 6}$ and $\la 3 \leftarrow 8\ra$. We reformulate the diagram in Figure \ref{fig:sp3_wall} to produce the diagram presented in Figure \ref{fig:sp3_junctions}. Vertices, edges and triangular faces correspond to vacua, walls and three-pronged junctions.

We study the three-pronged junction that divides vacua $\{\lra{1},\lra{2},\lra{3} \}$. The vacua are labelled by the rule (\ref{eq:spn_vac_label}):
\begin{align} \label{eq:spvac123}
&\lra{1} ~:~(\s_1+i\ta_1,\s_2+i\ta_2,\s_3+i\ta_3) \nn\\
&\quad\quad\quad =(m_1+in_1,m_2+in_2,m_3+in_3),  \nn\\
&\lra{2} ~:~(\s_1+i\ta_1,\s_2+i\ta_2,\s_3+i\ta_3) \nn\\
&\quad\quad\quad =(m_1+in_1,m_2+in_2,-m_3-in_3), \nn\\
&\lra{3} ~:~(\s_1+i\ta_1,\s_2+i\ta_2,\s_3+i\ta_3) \nn\\
&\quad\quad\quad =(m_1+in_1,-m_2-in_2,m_3+in_3).
\end{align}
We apply the worldvolume symmetry transformation (\ref{eq:equivcls}) to the moduli matrices of walls (\ref{eq:sp3_single_walls}) to produce the following moduli matrices $(h_{\sss{AI}}=\exp(a_{\sss{AI}}+ib_{\sss{AI}});~a_{\sss{AI}},b_{\sss{AI}} \in {\mathbf{R}})$:
\begin{align}
&H_{0\lra{1\leftrightarrow 2}}=
\lt(
\begin{array}{cccccc}
1    &   ~   &  ~        &   0        &    ~   &   ~    \\
~    &   1   &  ~        &   ~        &    0   &   ~   \\
~    &   ~   &  h_{33}   &   ~        &    ~   &   h_{36}     
\end{array}
\rt), \nn\\
&\quad \nn\\
&H_{0\lra{2\leftrightarrow 3}}=
\lt(
\begin{array}{cccccc}
1    &   ~        &  ~       &   0   &    ~   &   ~       \\
~    &   h_{22}   &  h_{23}  &   ~   &    0   &   ~       \\
~    &   ~        &  0       &   ~   &    -h_{23}   &  h_{22}            
\end{array}
\rt), \nn\\
&\quad \nn\\
&H_{0\lra{3\leftrightarrow 1}}=
\lt(
\begin{array}{cccccc}
1    &   ~       &  ~   &   0   &  ~       &   ~    \\
~    &   h_{22}  &  ~   &   ~   &  h_{25}  &   ~    \\
~    &   ~       &  1   &   ~   &  ~       &   0    
\end{array}
\rt).
\end{align}
As the moduli matrices have the worldvolume symmetry (\ref{eq:equivcls}), only one of $h_{\sss{AI}}$ parameters or the ratio of the two parameters in each moduli matrix is independent.

The wall solutions that interpolate $\lt\{\lra{1},\lra{2}\rt\}$, $\lt\{\lra{2},\lra{3}\rt\}$ and $\lt\{\lra{3},\lra{1}\rt\}$ are obtained by (\ref{eq:bps_sol_a}) with (\ref{eq:const_h1}) as follows:
\\~\\
$\noindent\bullet$$\Big\{\lra{1},\lra{2}\Big\}$
\begin{align} \label{eq:sp3_12_wall}
&\mca_{\lra{1\leftrightarrow 2}}=
\lt(
\begin{array}{cccccc}
\sqrt{c}    &   ~   &  ~        &   0        &    ~   &   ~    \\
~    &   \sqrt{c}    &  ~        &   ~        &    0   &   ~   \\
~    &   ~   &  \frac{p_{33}}{\sqrt{s_3}}   &   ~        &    ~   &   \frac{p_{36}}{\sqrt{s_3}}    
\end{array}
\rt), \nn\\
&p_{33}=\exp(m_3x^1+n_3x^2+a_{33}+ib_{33}), \nn\\
&p_{36}=\exp(-m_3x^1-n_3x^2+a_{36}+ib_{36}), \nn\\
&s_3=\frac{1}{c}\lt[\exp(2m_3x^1+2n_3x^2+2a_{33})+\exp(-2m_3x^1-2n_3x^2+2a_{36})\rt].
\end{align}
\\~\\
$\noindent\bullet$$\Big\{\lra{2},\lra{3}\Big\}$
\begin{align}\label{eq:sp3_23_wall}
&\mca_{\lra{2\leftrightarrow 3}}=
\lt(
\begin{array}{cccccc}
\sqrt{c}    &   ~        &  ~       &   0   &    ~   &   ~       \\
~    &   \frac{q_{22}}{\sqrt{t_2}}    &  \frac{q_{23}}{\sqrt{t_2}}  &   ~   &    0   &   ~       \\
~    &   ~        &  0       &   ~   &    \frac{q_{35}}{\sqrt{t_3}}   &  \frac{q_{36}}{\sqrt{t_3}}           
\end{array}
\rt), \nn\\
&q_{22}=\exp(m_2x^1+n_2x^2+a_{22}+ib_{22}), \nn\\
&q_{23}=\exp(m_3x^1+n_3x^2+a_{23}+ib_{23}), \nn\\
&q_{35}=\exp(-m_2x^1-n_2x^2+a_{23}+ib_{23}+i\pi),   \nn\\
&q_{36}=\exp(-m_3x^1-n_3x^2+a_{22}+ib_{22}), \nn\\
&t_2=\frac{1}{c}\lt[\exp(2m_2x^1+2n_2x^2+2a_{22})+\exp(2m_3x^1+2n_3x^2+2a_{23})\rt], \nn\\
&t_3=\frac{1}{c}\lt[\exp(-2m_2x^1-2n_2x^2+2a_{23})+\exp(-2m_3x^1-2n_3x^2+2a_{22})\rt].
\end{align}
\\~\\
$\noindent\bullet$$\Big\{\lra{3},\lra{1}\Big\}$
\begin{align} \label{eq:sp3_13_wall}
&\mca_{\lra{3\leftrightarrow 1}}=
\lt(
\begin{array}{cccccc}
\sqrt{c}    &   ~       &  ~   &   0   &  ~       &   ~    \\
~    &   \frac{r_{22}}{\sqrt{u_2}}  &  ~   &   ~   &  \frac{r_{25}}{\sqrt{u_2}}  &   ~    \\
~    &   ~       &  \sqrt{c}   &   ~   &  ~       &   0    
\end{array}
\rt), \nn\\
&r_{22}=\exp(m_2x^1+n_2x^2+a_{22}+ib_{22}), \nn\\
&r_{25}=\exp(-m_2x^1-n_2x^2+a_{25}+ib_{25}), \nn\\
&u_2=\frac{1}{c}\lt[\exp(2m_2x^1+2n_2x^2+2a_{22})+\exp(-2m_2x^1-2n_2x^2+2a_{25})\rt].
\end{align}
As before, only one of $a_{\sss{AI}}$ parameters is independent in each wall solution.

The position of wall $\lra{1 \leftrightarrow 2}$ is $\mathrm{Re}(p_{33})=\mathrm{Re}(p_{36})$:
\begin{align} \label{eq:sp_wall_12_pos}
2m_3x^1+2n_3x^2+(a_{33}-a_{36})=0.
\end{align}
The position of wall $\lra{2 \leftrightarrow 3}$ is $\mathrm{Re}(q_{22})=\mathrm{Re}(q_{23})$ and $\mathrm{Re}(q_{35})=\mathrm{Re}(q_{36})$:
\begin{align} \label{eq:sp_wall_23_pos}
(m_2-m_3)x^1+(n_2-n_3)x^2+(a_{22}-a_{23})=0.
\end{align}
The position of wall $\lra{3 \leftrightarrow 1}$ is $\mathrm{Re}(r_{22})=\mathrm{Re}(r_{25})$:
\begin{align} \label{eq:sp_wall_13_pos}
2m_2x^1+2n_2x^2+(a_{22}-a_{25})=0.
\end{align}
There is a consistency condition:
\begin{align}
(a_{22}-a_{25})-(a_{33}-a_{36})-2(a_{22}-a_{23})=0.
\end{align}
Therefore, there are two independent $a_{\sss{AI}}$ parameters to describe the junction position as expected.
We can compute the position of the junction from (\ref{eq:sp_wall_12_pos}), (\ref{eq:sp_wall_23_pos}) and (\ref{eq:sp_wall_13_pos}). The position of the junction that divides the vacua $\{\lra{1},\lra{2},\lra{3}\}$ in (\ref{eq:spvac123}) is
\begin{align}
(x,y)=\lt(\frac{T_1}{T_3},\frac{T_2}{T_3}\rt),
\end{align}
\begin{align}
&T_1=2(-a_{33}+a_{36})n_2+2(a_{22}-a_{25})n_3, \nn\\
&T_2=2(a_{33}-a_{36})m_2+2(-a_{22}+a_{25})m_3, \nn\\
&T_3=-4m_2 n_3 + 4m_3 n_2.
\end{align}
%

%%%%%%%%%%%%%%%%%%%%%%%%%%%%%%%%%%%%%%%%%%%%%%%%%%%%%%%%%%%%%%%%%%%%%%%%%%%%
\section{Summary} \label{sec:summary}
\setcounter{equation}{0}
%%%%%%%%%%%%%%%%%%%%%%%%%%%%%%%%%%%%%%%%%%%%%%%%%%%%%%%%%%%%%%%%%%%%%%%%%%%

We have proposed on-shell ${\mathcal{N}}=2$ nonlinear sigma models on $SO(2N)/U(N)$ and $Sp(N)/U(N)$ by holomorphically embedding the models in the hyper-K\"{a}hler nonlinear sigma model on the cotangent bundle of the Grassmann manifold $T^\ast G_{2N,N}$ in the ${\mathcal{N}}=1$ superspace formalism.

We have applied the moduli matrix formalism to three-pronged junctions of the mass-deformed nonlinear sigma models on $SO(2N)/U(N)$ and $Sp(N)/U(N)$ by making use of the pictorial representation \cite{Lee:2017kaj,Arai:2018tkf} and the diagram method \cite{Eto:2005cp,Shin:2018chr}.  We have presented three-pronged junctions and have calculated positions of junctions of the mass-deformed nonlinear sigma models on $SO(8)/U(4)$ and $Sp(3)/U(3)$, which are non-Abelian gauge theories.

It is shown in \cite{Lee:2017kaj,Arai:2018tkf} that we can produce the whole structures of vacua and walls of the mass-deformed nonlinear sigma models on $SO(2N)/U(N)$ and $Sp(N)/U(N)$ from the vacuum structures that are connected to the maximum number of elementary walls for generic $N$. We have observed in this paper that we can construct three-pronged junctions and we can find positions of any three-pronged junctions of the mass-deformed nonlinear sigma models on $SO(8)/U(4)$ and $Sp(3)/U(3)$. Therefore, this approach can be applied to the mass-deformed nonlinear sigma models on $SO(2N)/U(N)$ and $Sp(N)/U(N)$ with generic $N$. As a sequel to this paper, we have applied the approach to the mass-deformed nonlinear sigma models on $SO(2N)/U(N)$ and $Sp(N)/U(N)$ and have discussed vacua, walls and three-pronged junctions for generic $N$ \cite{Kim:2020obf}.

~~\\~~\\
\noindent {\bf Acknowledgement}

We would like to thank M. Arai for his early participation. We are indebted to M. Eto, S. Fedoruk, E. Ivanov, S. Kuzenko and M. Nitta for valuable comments and suggestions. S. S. is grateful to the organisers of the International Workshop Supersymmetries and Quantum Symmetries - SQS'19, where the main part of this work was done. T. K. was supported by the National Research Foundation of Korea (NRF) grant funded by the Korea government (MEST)  (NRF-2018R1D1A1B07051127, NRF-2019R1A6A1A10073079). S. S. was supported by the National Research Foundation of Korea (NRF) grant funded by the Korea government (NRF-2017R1D1A1B03034222, NRF-2021R1A2C1009197).

%%%%%%%%%%%%%%%%%%%%%%%%%%%%%%%%%%%%%%%%%%%%%%%%%%%%
%
% Appendix
%
%%%%%%%%%%%%%%%%%%%%%%%%%%%%%%%%%%%%%%%%%%%%%%%%%%%%
\appendix
\def\thesection{Appendix \Alph{section}}
\setcounter{equation}{0}
\renewcommand{\theequation}{\Alph{section}.\arabic{equation}}

%%%%%%%%%%%%%%%%%%%%%%%%%%%%%%%%%%%%%%%%%%%%%%%%%%%%%%%%%%%%%%%%%%%%%%%%%%%%
\section{On-shell ${\mathcal{N}}=2$ supersymmetry of the action (\ref{ea:n=2nlsmord}) }  \label{app:eq}
\setcounter{equation}{0}
%%%%%%%%%%%%%%%%%%%%%%%%%%%%%%%%%%%%%%%%%%%%%%%%%%%%%%%%%%%%%%%%%%%%%%%%%%%%
In this Appendix, we present the relation between the component field action of (\ref{ea:n=2nlsmord}) \cite{Arai:2002xa,Arai:2003tc} and the component field action of the mass-deformed nonlinear sigma model on $T^\ast G_{N+M,M}$ that is constructed in the harmonic superspace formalism \cite{Galperin:2001uw}. It is discussed in detail in \cite{Kim:2020obf}.

We follow the convention of \cite{Galperin:2001uw}. We repeat the action (\ref{ea:n=2nlsmord}) with component fields:
\begin{align}\label{ea:n=2nlsmordb}
S=&\int d^4x \Big\{\int d^4\th \tr\Big(\Ph\bar{\Ph} e^V+\bar{\Psi}\Psi e^{-V}-c V \Big) \nn \\
&+\Big[\int d^2\th \tr\Big(\Xi(\Ph\Psi-b I_M)+\Ph M\Psi\Big)+\mathrm{(c.t.)}\Big]\Big\} ,\nn\\
&(c\in \mathbf{R}_{\geq0},~ b\in \mathbf{C}),
\end{align}
\begin{align}\label{eq:n=1_formalism}
&\Ph_{\sss{A}}^{~\sss{I}}(y)={\mca}_{\sss{A}}^{~\sss{I}}(y)+\sqrt{2}\th\z_{\sss{A}}^{~\sss{I}}(y)+\th\th F_{\sss{A}}^{~\sss{I}}(y), \quad (y^\m=x^\m+i\th\s^\m\bar{\th}),  \nn\\
&\Psi_{\sss{I}}^{~\sss{A}}(y)={\mcb}_{\sss{I}}^{~\sss{A}}(y)+\sqrt{2}\th\e_{\sss{I}}^{~\sss{A}}(y)+\th\th G_{\sss{I}}^{~\sss{A}}(y) , \nn\\
&V_{\sss{A}}^{~\sss{B}}(x)=2\th\s^\m\bar{\th}A_{\m{\sss{A}}}^{~~\sss{B}}(x)+ i\th\th\bar{\th}\bar{\l}_{\sss{A}}^{~\sss{B}}(x)-i\bar{\th}\bar{\th}\th\l_{\sss{A}}^{~\sss{B}}(x)+\th\th\bar{\th}\bar{\th}D_{\sss{A}}^{~\sss{B}}(x), \nn\\
&\Xi_{\sss{A}}^{~\sss{B}}(y)=-{\mcs}_{\sss{A}}^{~\sss{B}}(y)+\th\o_{\sss{A}}^{~\sss{B}}(y)+\th\th K_{\sss{A}}^{~\sss{B}}(y), \nn\\
& (i=1,\cdots,N+M;~ A=1,\cdots,M).
\end{align}
By solving the auxiliary fields, we obtain the component field action:
\begin{align} \label{eq:n=1_component_act}
S=\int & d^4x \tr\Big[ D_\m\mca\overline{D^\m\mca}
+\overline{D_\m\mcb}D^\m\mcb  \nn\\
&-\frac{i}{2}\z\s^\m \overline{D_\m\z} + \frac{i}{2} D_\m\z\s^\m\bar{\z}
-\frac{i}{2}\bar{\e} \bar{\s}^\m D_\m \e + \frac{i}{2} \overline{D_\m\e}\bar{\s}^\m\e \nn\\
&-\frac{\sqrt{2}i}{2}\lt(\mca\bar{\z}\bar{\l}-\l\z\bar{\mca}\rt) 
-\frac{\sqrt{2}}{2}\lt(\o\mca\e+\bar{\e}\bar{\mca}\bar{\o}\rt)\nn\\
&-\frac{\sqrt{2}i}{2}\lt(\bar{\mcb}\e\l-\bar{\l}\bar{\e}\mcb\rt)
-\frac{\sqrt{2}}{2}\lt(\bar{\mcb}\bar{\z}\bar{\o}+\o\z\mcb\rt) \nn\\
&+\mcs\z\e+\bar{\e}\bar{\z}\bar{\mcs}-\z M \e - \bar{\e}\bar{M}\bar{\z} \nn\\
&-|\mca M -\mcs\mca|^2 - |\bar{\mcb}\bar{M}-\bar{\mcs}\bar{\mcb}|^2  \nn\\
&+(\mca\bar{\mca}- \bar{\mcb}\mcb-cI_M)D+K(\mca\mcb-bI_M)+(\bar{\mcb}\bar{\mca}-b^\ast I_M)\bar{K}
\Big],
\end{align}
\begin{align}\label{eq:n=1_covariant_der}
&D_\m\mca=\p_\m\mca-iA_\m \mca, \quad \overline{D_\m\mcb}=\p_\m\bar{\mcb}-i A_\m \bar{\mcb}  , \nn\\
&D_\m\z=\p_\m\z-iA_\m \z, \quad ~~ \,  \overline{D_\m\e}=\p_\m \bar{\e} - i  A_\m \bar{\e}.
\end{align}

The mass-deformed nonlinear sigma model on $T^\ast G_{N+M,M}$ that is constructed in the harmonic superspace formalism \cite{Galperin:2001uw} is
\begin{align} \label{eq:hssfldact}
&{\mathcal{S}}=-\int
d\zeta^{(-4)}du\mathrm{Tr}\Big[\widetilde{\ph^+}(D^{++}+iV^{++})\ph^++\xi^{++}V^{++}\Big], \\
&d\zeta^{(-4)}=d^4x d^2\th^+ d^2\bar{\th}^+, \nn
\end{align}
\begin{align}\label{eq:ham_fld_op}
&\ph^+(\z,u)=F^+(x,u)+\sqrt{2}\th^{+\a}\psi_\a(x,u)+\sqrt{2}\bar{\theta}^+_{\dot{\a}}\bar{\varphi}^{\dot{\a}}(x,u) \nn\\
&\quad\quad\quad\quad+(\th^+)^2M^-(x,u)+(\bar{\th}^+)^2N^-(x,u)+i\th^+\s^\m\bar{\th}^+A_\m^- (x,u)  \nn\\
&\quad\quad\quad\quad  
+\sqrt{2}(\th^+)^2\bar{\th}^+_{\dot{\a}}\bar{\chi}^{(-2)\dot{\a}}(x,u)+\sqrt{2}(\bar{\th}^+)^2\th^{+\a}\ta^{(-2)}_{\a}(x,u)\nn\\
&\quad\quad\quad\quad +(\th^+)^2(\bar{\th}^+)^2 D^{(-3)}(x,u),  \nn\\
&V^{++}(\z,u)=-i(\th^+)^2\S(x)+i(\bar{\th}^+)^2\bar{\S}(x)+2i\th^+\s^\m \bar{\th}^+V_\m(x) \nn\\
&\quad\quad\quad\quad ~ ~ +\sqrt{2}(\bar{\th}^+)^2\th^{+\a}\xi^{i}_{{\sss{V}}\a}(x)u_i^{-}-\sqrt{2}(\th^+)^2\bar{\th}^+_{\dot{\a}}\bar{\xi}^{\dot{\a}i}_{\sss{V}}(x)u_i^{-} \nn\\
&\quad\quad\quad\quad ~ ~+3\th^{+2}\bar{\th}^{+2}D_{\sss{V}}^{ij}(x)u_i^- u_j^-, \nn \\
&D^{++}=\p^{++}-2i\th^+\s^\m\bar{\th}^+\p_\m+i(\th^+)^2(\p_5-i\p_6)-i({\bar{\th}}^+)^2(\p_5+i\p_6)\nn\\
&\quad\quad~ +\th^{+\a}\frac{\p}{\p\th^{-\a}}+\bar{\th}^{+\dot{\a}}\frac{\p}{\p\bar{\th}^{-\dot{\a}}}.
\end{align}
We solve the kinematic equations that eliminate the infinite sets of auxiliary fields in the harmonic expansions. For example, the equation $\p^{++}F^+=0$ is solved by $F^+(x,u)=f^i(x)u_i^+$. The detailed calculation is presented in \cite{Kim:2020obf}. The component field action is
\begin{align}\label{ea:harmonic_gr}
{\mathcal{S}}=\int d^4x &\tr\Big[\overline{D_\m f_i} D^\m f^i -i\bar{\psi}\bar{\s}^\m D_\m\psi
-i\varphi \s^\m \overline{D_\m {\varphi}} \nn\\
&+\frac{i}{2}\bar{f}_i\xi_{\sss{V}}^i\psi -\frac{i}{2}\bar{f}_i\bar{\xi}_{\sss{V}}^i\bar{\varphi}
-\frac{i}{2}\bar{\psi}\bar{\xi}_{\sss{V}i}f^i -\frac{i}{2}\varphi\xi_{\sss{V}i}f^i  \nn\\
&-\frac{1}{2}|f^i\mcm-\S f^i|^2 -\frac{1}{2}|f^i\bar{\mcm}-\bar{\S}f^i|^2 \nn\\
&-\bar{\psi}\lt(\bar{\varphi}\mcm-\S\bar{\varphi}\rt)-\varphi\lt(\psi\bar{\mcm}-\bar{\S}\psi\rt) \nn\\
&+if^{(i}\bar{f}^{j)}D_{V(ij)}- \xi^{(ij)}D_{\sss{V}(ij)}\Big],
\end{align}
\begin{align}
&D_\m f^i=\p_\m f^i  -iV_\m f^i, \nn\\
&D_\m\psi=\p_\m \psi  -iV_\m \psi, \quad
\overline{D_\m\varphi}= \p_\m \bar{\varphi}  -iV_\m  \bar{\varphi}.
\end{align}

The action (\ref{eq:n=1_component_act}) and the action (\ref{ea:harmonic_gr}) are equivalent. The relations between the component fields are presented below:
\begin{align} \label{eq:comp_cmpr}
&f^1=\mca,\quad f^2=\bar{\mcb},\quad \psi=\z,\quad \bar{\varphi}=-\bar{\e},  \nn \\
&V_\mu=A_\mu,\quad \xi_{\sss{V}}^1=\sqrt{2}\l, \quad \xi_{\sss{V}}^2=\sqrt{2}i\o, \nn\\
&\S:=\frac{e^{i\a}}{\sqrt{2}}\lt(\S\bar{\S}+\bar{\S} \S \rt)^{\frac{1}{2}}, \quad \S=-\bar{\mcs}, \nn\\
&\bar{\S}:=\frac{e^{-i\a}}{\sqrt{2}}\lt(\S \bar{\S}+\bar{\S}\S\rt)^{\frac{1}{2}}, \quad \bar{\S}= -\mcs, \nn \\
&\mcm:=\frac{e^{i\b}}{\sqrt{2}}\lt(\mcm\bar{\mcm}+\bar{\mcm}\mcm\rt)^{\frac{1}{2}}, \quad \mcm=-\bar{M},\nn\\
&\bar{\mcm}:=\frac{e^{-i\b}}{\sqrt{2}}\lt(\mcm\bar{\mcm}+\bar{\mcm}\mcm\rt)^{\frac{1}{2}}, \quad \bar{\mcm}=-M, \nn\\
&D_{\sss{V}11}=iK,\quad D_{\sss{V}22}=-i\bar{K},\quad D_{\sss{V}(12)}=-iD, \nn \\
&\xi^{11}=-ib,\quad \xi^{22}=ib^\ast,\quad \xi^{(12)}=\frac{ic}{2}. 
\end{align}

\end{document}